\documentclass[
 reprint,
 nofootinbib,
 amsmath,
 amssymb,
 aps,
]{revtex4-2}
\usepackage{ulem}
\usepackage{graphicx}
\usepackage[utf8]{inputenc}
\usepackage{verbatim}
\usepackage{dcolumn}
\usepackage{bm}
\usepackage{hyperref}
\usepackage{bookmark}
\usepackage[all]{hypcap}
\usepackage{makecell}  
\usepackage{xspace}
\usepackage[usenames,dvipsnames,svgnames]{xcolor}
\usepackage{subcaption}
\usepackage{ragged2e}
\usepackage{booktabs}
\newcommand{\jcaption}[1]{\caption{\justifying #1}}
\definecolor{darkgreen}{cmyk}{0.9,0,0.9,0.6}
\definecolor{brickred}{RGB}{210,65,84}
\hypersetup{colorlinks=true,
	citecolor=Blue,
	linkcolor=Magenta,
    	filecolor=NavyBlue,
	urlcolor=NavyBlue}

\graphicspath{{figure/}}

\def\be{\begin{equation}}
\def\ee{\end{equation}}
\def\ba{\begin{eqnarray}}
\def\ea{\end{eqnarray}}

\newcommand{\aap}{Astronomy \& Astrophysics\xspace}
\newcommand{\apjl}{Astrophysical Journal Letters\xspace}

\def\({\left(}
\def\){\right)}
\def\[{\left[}
\def\]{\right]}

\usepackage{cleveref}
\usepackage[labelstoglobalaux]{bibunits}
\begin{document}

\title{Effects of dark dipole radiation on eccentric supermassive black hole binary inspirals}

\author{Mu-Chun Chen$^{a}$}
\email{chenmuchun23@mails.ucas.ac.cn}

\author{Yan Cao$^{b}$}
\email{yancao@smail.nju.edu.cn}

\affiliation{\begin{footnotesize}
		${}^a$School of Astronomy and Space Science, University of Chinese Academy of Sciences (UCAS), Beijing 100049, China\\
        ${}^b$School of Physics, Nanjing University, Nanjing 210093, China\\
		\end{footnotesize}}

\begin{abstract}
The final-parsec problem has long posed a central challenge in understanding the merger of supermassive black hole binaries. In this paper, we investigate a scenario in which a dark scalar or vector field is sourced by eccentric binaries, leading to accelerated mergers through additional dipole radiation, and thereby extending the range of masses for which the binary can merge within a Hubble time. The Newtonian-order radiation fluxes from an eccentric charged Keplerian binary are derived using general results for localized periodic sources in flat spacetime. We find that dipole radiation, although insufficient to fully resolve the final-parsec problem, can alter the low-frequency spectrum of the stochastic gravitational wave background from supermassive black hole binary inspirals. We construct a simplified model for the spectrum and perform a Bayesian analysis using the current pulsar timing array data.
\end{abstract}

\maketitle

\tableofcontents

\onecolumngrid

\section{Introduction}
\label{sec:intro}
The final-parsec problem refers to a potential stall in the hardening of supermassive black hole binaries (SMBHBs) at semimajor axes $a\!\lesssim\!1\,\mathrm{pc}$ before gravitational wave (GW) radiation dominates the orbital evolution. In nearly spherical, collisionless galactic nuclei, interactions with stars exhaust the phase-space region of low-angular-momentum (loss-cone) orbits that intersect the binary. Because relaxation-driven refilling is inefficient in this geometry, the scattering-induced hardening rate may become too small for the binary to bridge the gap to the GW-dominated regime within a Hubble time~\cite{1980Natur.287..307B, Quinlan_1996, Merritt_2005, Vasiliev_2015}. A variety of astrophysical mechanisms have been proposed to address this challenge, introducing additional dynamical dissipation. These include triaxial potentials to enhance loss-cone replenishment, three-body scattering with stars or dark matter~\cite{2004ApJ...606..788M, Berczik_2006, Vasiliev_2015, Gualandris_2016}, and dynamical friction to accelerate binary mergers~\cite{Chen:2025jch, bromley2023supermassiveblackholebinaries, koo2024finalparsecproblemblack, Shen:2023pan, Fischer_2024, Hu_2025, Ding:2025nxe}.

In addition to environmental effects, the intrinsic dynamics of SMBHBs may also be influenced by the physics of hidden sectors, one possibility being that they effectively carry dark scalar or vector charges. These charges may be associated with particles beyond the Standard Model~\cite{DERUJULA1990173,Holdom:1985ag,Damour_2010,Wong_2019,PhysRevD.102.023005,Dror_2021}, and they may also arise in modified theories of gravity~\cite{PhysRevLett.120.131103,PhysRevLett.120.131104,Seymour_2020,Hu_2024}. In general relativity, black holes can carry multiple U(1) charges and may also support non-Abelian hair~\cite{PhysRevD.12.2212,herdeiro2017reissnernordstromblackholesnonabelian,G_mez_2023}. While the no-hair theorem precludes black holes with stable scalar or Proca charges, the timescale for discharge of a Proca field can be sufficiently long when the field is extremely light~\cite{Coleman_1992}. Furthermore, the BH may be surrounded by charged matter, even if it is not charged itself. So long as the size of the bound charge cloud is much smaller than the Compton wavelength of the field quanta, the dressed BH can be effectively treated as a point charge. The dynamics of charged compact binaries have been investigated by numerous works, both for the massless fields~\cite{Cardoso_2016,Khalil_2018,Kuntz_2019,Patil_2020,Liu_2020,Christiansen_2021,Cardoso_2021,gupta2022binarydynamicseinsteinmaxwelltheory,Henry_2023,Henry_2024} and the massive fields~\cite{LI198620,PhysRevD.49.6892,Huang:2018pbu,Bhattacharyya:2023kbh,Diedrichs:2023foj,owen2025constrainingdarksectoreffectsusing}, with broad theoretical implications~\cite{Mohanty_1996,Croon_2018,Alexander_2018,KumarPoddar:2019ceq,PhysRevD.102.023005,Seymour_2020,Gupta_2021,Zhang_2021,Chen:2024ery,Liu:2024qzp,Liu:2025zuz,Bai:2024pki,Bhalla_2024,wang2025tinyhalosstirspacetime}. At the Newtonian order,\footnote{A post-Newtonian (PN) order of $n$ refers to the correction that scales with $v^{2n}$ relative to the leading term, where $v$ is the orbital velocity. In the presence of charges, we also refer to the leading contribution to the scalar or vector radiation as the Newtonian-order (0PN) term, although the 0PN dipole radiation fluxes of a massless field scales with $v^{-2}$ relative to the 0PN fluxes of gravitational radiation (see Sec.~\ref{sec:Charged binary}).} the effects of dark charges include a Yukawa force and dipole radiation, if the dipole moment is not suppressed by the difference in charge-to-mass ratios. If the scalar or vector field is sufficiently light or if the Yukawa force is negligible, the binary's orbital motion is still approximately Keplerian. For an eccentric Keplerian binary, the Newtonian-order energy fluxes of massive scalar and vector dipole radiation were derived in \cite{PhysRevD.49.6892}, while the corresponding angular momentum fluxes were obtained in \cite{Cheng:2023qys}. The general expressions for the energy fluxes of massive scalar and vector radiation from a localized periodic source in flat spacetime were derived in \cite{PhysRevD.49.6892}; they can also be obtained using a momentum-space approach~\cite{Mohanty_1996,KumarPoddar:2019ceq}. In the massless case, the Newtonian-order fluxes from a charged eccentric Keplerian binary were derived in \cite{Cardoso_2016,Liu_2020,Christiansen_2021}.


Over the past two years, pulsar timing arrays (PTAs) have reported compelling evidence for a nanohertz stochastic gravitational wave background (SGWB), including the emergence of the Hellings-Downs inter-pulsar correlation expected for an unpolarized and isotropic SGWB~\cite{2023ApJ...951L...8A,2023ApJ...951L...6R,2023AA...678A..50E,2023RAA....23g5024X}. Independent analyses from NANOGrav (15-year)~\cite{PhysRevD.109.103012,2023ApJ...952L..37A}, EPTA + InPTA (DR2)~\cite{Antoniadis_2022}, and PPTA (DR3)~\cite{Zic_2023} all point to a characteristic strain with an amplitude of $\mathcal{O}(10^{-14})$ at $f=1\,\mathrm{yr}^{-1}$, broadly consistent with a population of inspiraling SMBHBs. It is also observed that the characteristic strain spectrum shows a flattening trend in the nHz range. These findings provide new observational insights into the processes of galaxy assembly and black hole-galaxy coevolution, while also motivating refined modeling of the background’s spectral shape and anisotropy, including potential interpretations beyond general relativity~\cite{2023ApJ...951L..11A,PhysRevLett.132.171002,PhysRevD.109.L061502,Smarra_2023}, and involving additional dissipation processes in the SMBHB inspirals beyond gravitational radiation~\cite{Aghaie_2024,guo2025ultralightbosonionizationcomparablemass, chen2025galaxytomographygravitationalwave, Chen:2025jch, alonsoálvarez2024selfinteractingdarkmattersolves, Shen_2025, Hu_2025, Ding:2025nxe}.

Observations of the SGWB by PTAs may also probe the possible dark charge content of SMBHs through the resulting modifications to orbital evolution, which have been analyzed in \cite{Dror_2021} for circular SMBHB inspirals. However, since the oscillatory components of gravitational waves in the nHz band are generated by SMBHBs long before their mergers, a significant fraction of the contributing binaries may retain high eccentricities. The dark radiation from charged SMBHBs will also directly contribute to the SGWB if the bosonic fields correspond to additional gravitational degrees of freedom~\cite{Zhang:2023lzt}.

In this paper, we study the effects of dipole radiation sourced by dark scalar or vector charges on the orbital evolution of eccentric SMBHBs, its possible role in alleviating the final-parsec problem, and its observational signatures in the SGWB from SMBHB inspirals. Compared with circular binaries, dipole radiation begins to accelerate the inspiral of eccentric binaries at lower orbital frequencies. We derive general expressions for the energy, linear momentum and angular momentum fluxes of massive scalar and vector dipole radiation from a localized periodic source in flat spacetime and use them to obtain the Newtonian-order radiation fluxes from an eccentric charged binary in the Keplerian approximation. Incorporating these fluxes into the adiabatic evolution of orbital eccentricity and semimajor axis, we compute the characteristic strain spectrum for a representative population of SMBHBs under simplified assumptions. We then confront this model with PTA free-spectrum data through a Bayesian analysis.

The remainder of this paper is organized as follows. In Sec.~\ref{sec.2}, we derive the dipole radiation fluxes of massive scalar and vector fields from an eccentric charged Keplerian binary. The secular orbital evolution in the presence of dipole radiation is examined in Sec.~\ref{sec.3}. In Sec.~\ref{sec.4}, we construct the SGWB using a population model and perform a Bayesian analysis using current PTA data. We briefly summarize the results and present our conclusions in Sec.~\ref{sec:conclusion}. Throughout this paper, we adopt the flat spacetime metric $\eta_{ab} = \mathrm{diag}(1,-1,-1,-1)$ and natural units with $\hbar = c = G = 1$.

\section{Dipole radiation from an eccentric charged Keplerian binary}
\label{sec.2}
In general relativity, the action for two point-like bodies carrying scalar and vector charges, sourcing a real scalar field with mass $m_\phi$ and a real vector field with mass $m_A$ minimally coupled to gravity, is given by
\begin{equation}\label{action}
\begin{aligned}
S =&\sum_{I=1,2} \left[q_{IA}\int dX^a_I A_a +\int (-m_I+q_{I\phi}\,\phi)\sqrt{g_{ab}dX_I^adX_I^b}\right]
\\
&+\int d^4x\,\sqrt{-g}\left(-\frac{R}{16\pi}+\frac{1}{2}\partial_a\phi\,\partial^a\phi-\frac{1}{2}m_\phi^2\phi^2-\frac{1}{4}F_{ab}F^{ab}+\frac{1}{2}m_A^2A_aA^a\right);
\end{aligned}
\end{equation}
where the spacetime coordinates of the $I$-th body are denoted by $X_I^a=(t,\mathbf{X}_I(t))$; its mass, scalar, and vector charges are $\{m_I,q_{I\phi},q_{IA}\}$. Here, we neglect possible higher-order couplings to the worldlines and nongravitational interactions of the bosonic fields. We assume that the massive vector field is described by the Proca Lagrangian; the Maxwell field is recovered in the massless limit $m_A\to 0$. For example, a Kerr-Newman black hole with mass $m_\text{BH}$, electric charge $q_A$, and spin parameter $a$ has $q_A^2<4\pi(m_\text{BH}^2-a^2)$, with $m_A=0$. The possible superradiant production from rotating BHs is neglected since we focus on extremely light fields, for which the superradiant instability (if present) develops on timescales much longer than relevant astrophysical timescales.\footnote{The superradiant instability is absent for a free massless bosonic field. For a free massive bosonic field, it is suppressed if $m\,m_\text{BH}$ is very small, see, for example, Sec.~4.6.5 of \cite{book_superradiance}. The instability rate of the fastest growing mode (for dimensionless BH spin $\chi=a/m_\text{BH}\gg m\,m_\text{BH}$) is $\omega_I\approx \frac{1}{48}\frac{\chi}{1+\sqrt{1-\chi^2}}(m\,m_\text{BH})^8m$ in the scalar case and $\omega_I\approx 2\frac{\chi}{1+\sqrt{1-\chi^2}}(m\,m_\text{BH})^6m$ in the vector case. If $m\lesssim 10^{-24}\,\text{eV}$, even for $m_\text{BH}=10^{10}M_\odot$, then $m\,m_\text{BH}\lesssim 8\times10^{-5}$; the corresponding instability timescale for $\chi=1$ is $\tau_I=1/(2\omega_I)\gtrsim 10^{35}\,\text{yr}$ in the scalar case and $\tau_I\gtrsim 10^{25}\,\text{yr}$ in the vector case, vastly exceeding the age of the Universe ($\sim 10^{10}\,\text{yr}$).} We also neglect the background of the scalar or vector field, which could interact with the bodies through direct coupling and gravitational effects~\cite{Nacir_2018, Blas_2020,bromley2023supermassiveblackholebinaries,koo2024finalparsecproblemblack,PhysRevLett.132.211401,tomaselli2025scatteringwavedarkmatter}. For the binary, we define $\mathbf{X}\equiv \mathbf{X}_1-\mathbf{X}_2$, $r\equiv |\mathbf{X}|$, $\mathbf{v}\equiv \dot{\mathbf{X}}\equiv \frac{\mathrm{d}\mathbf{X}}{\mathrm{d}t}$, and $v\equiv|\mathbf{v}|$.

\subsection{Radiation from a localized periodic source}
At leading post-Newtonian (PN) order, the potential modes of the graviton, scalar, and vector fields do not affect the scalar and vector radiation, which are therefore described by the wave equations in flat spacetime,
\begin{align}
(\square-m_A^2) A^a &=J^a(t,\mathbf{x})=\sum_I q_{IA}\, v^a_I\,\delta^3(\mathbf{x}-\mathbf{X}_I(t));\label{Proca_equ}
\\
(\square-m_\phi^2) \phi &=-n(t,\mathbf{x})=-\sum_I q_{I\phi}\, \delta^3(\mathbf{x}-\mathbf{X}_I(t));
\end{align}
with $v^a_I=(1,\mathbf{v}_I)$ and $\square \equiv -\partial_t^2+\partial_i \partial_i$. For simplicity, we assume that the scalar and vector charges are time-independent.\footnote{For time-dependent scalar charges, the monopole source term $\int d^3x\,e^{-i\mathbf{k}\cdot\mathbf{x}}\left[\int dt\,n(t,\mathbf{x})\,e^{i\omega t}\right]\supset \int d^3x\left[\int dt\,n(t,\mathbf{x})\,e^{i\omega t}\right]=\int dt\left(\sum_I q_{I\phi}\right)e^{i\omega t}$ generally does not vanish for $\omega\ne 0$, and the resulting monopole radiation depends on the specific time evolution of the charges.} For the vector charge, this implies $\partial_a J^a=0$ such that $\partial_a A^a=\dot A_0-\partial_i A_i=0$. We consider the scalar and vector cases separately, so the subscripts $\{\phi,A\}$ will be omitted in each case when there is no ambiguity. The solution of Eq.~\eqref{Proca_equ} in momentum space is
\begin{equation}
A^a(\omega,\mathbf{k})= \frac{J^a(\omega,\mathbf{k})}{\omega^2-|\mathbf{k}|^2-m^2};
\end{equation}
with $J^a(\omega,\mathbf{k})$ being the Fourier transform of the source term
\begin{align}
	J^a(\omega,\mathbf{k}) &=\int d^3x\,e^{-i\mathbf{k}\cdot\mathbf{x}}\left[\int_{-\infty}^\infty dt\,J^a(t,\mathbf{x})\,e^{i\omega t}\right]
	.
\end{align}
For a localized source, the on-shell radiation field in $A^a$ is given by~\cite{LI198620,PhysRevD.49.6892}
\begin{align}\label{radiation_field}
    A^a(t,\mathbf{x}=r\mathbf{n})=-\frac{1}{4\pi r}\left\{\int_{-\infty}^{-m} \frac{d\omega}{2\pi} J^a(\omega,\mathbf{k})\,e^{i\omega\left[\sqrt{1-(m/\omega)^2}\mathbf{n}\cdot\mathbf{x}- t\right]}
    +
    \int_{m}^\infty \frac{d\omega}{2\pi} J^a(\omega,\mathbf{k})\,e^{i\omega\left[\sqrt{1-(m/\omega)^2}\mathbf{n}\cdot\mathbf{x}- t\right]}
    \right\}
.
\end{align}

We consider a source with a temporal period $T=2\pi/\Omega$, for which
\begin{align}
J^a(\omega,\mathbf{k}) &=\sum_{n} 2\pi \delta(\omega-\Omega_n)\,J^a_n(\mathbf{k});
\\
J^a_n(\mathbf{k}) &=\int d^3x\,e^{-i\mathbf{k}\cdot\mathbf{x}}\left[\frac{1}{T}\int_0^Tdt\,J^a(t,\mathbf{x})\,e^{i\Omega_n t}\right]; \label{J^a_n}
\end{align}
where $\Omega_n\equiv n\Omega$ and $n\in\mathbb{Z}$, thus Eq.~\eqref{radiation_field} becomes
\begin{align}
    A^a(t,\mathbf{x}=r\mathbf{n})&=-\frac{1}{4\pi r}\sum_{|n|\ge n_0}J^a_{(n)}\,e^{i\left[\mathbf{k}^{(n)}\cdot\mathbf{x}-\Omega_nt\right]}
	,\label{vector_radiation_field}
\end{align}
with $J^a_{(n)} = J^a_n(\mathbf{k}^{(n)})$, $\mathbf{k}^{(n)}=k_n\mathbf{n}\equiv\Omega_n\sqrt{1-(m/\Omega_n)^2}\,\mathbf{n}$, and $ n_0\equiv m/\Omega$. Note that $J_{-n}^a=J_n^{a*}$, and $J_{(n)}^0=\mathbf{J}_{(n)}\cdot\mathbf{k}^{(n)}/\Omega_n$. Similarly, the scalar radiation field from a localized periodic source is
\begin{align}
	\phi(t,\mathbf{x}=r\mathbf{n})&=\frac{1}{4\pi r}\sum_{|n|\ge n_0}n_{(n)}\,e^{i\left[\mathbf{k}^{(n)}\cdot\mathbf{x}-\Omega_nt\right]}, \label{scalar_radiation_field}
    \\
    n_{(n)} & = \int d^3x\,e^{-i\mathbf{k}^{(n)}\cdot\mathbf{x}}\left[\frac{1}{T}\int_0^Tdt\,n(t,\mathbf{x})\,e^{i\Omega_n t}\right].\label{n_{(n)}}
\end{align}

The time-averaged radiation fluxes of energy, linear momentum, and angular momentum can be obtained as
\begin{align}
P=r^2\int d\Omega_\mathbf{n}\,\sum_{n \ge n_0} \rho^{(n)}\,v_g^{(n)},
\quad
\mathcal{F}_i= r^2\int d\Omega_\mathbf{n}\,\sum_{n \ge n_0} p^{(n)}_i\,v_g^{(n)},
\quad
\tau_i = r^2\int d\Omega_\mathbf{n}\,\sum_{n \ge n_0} j^{(n)}_i\,v_g^{(n)},
\end{align}
where $d\Omega_\mathbf{n}$ is the element of the solid angle, and $\{v_g^{(n)},\rho^{(n)}, \mathbf{p}^{(n)}, \mathbf{j}^{(n)}\}$ are the group velocity, time-averaged volume density of energy, linear momentum, and angular momentum of the outgoing $\mathbf{k}^{(n)}$-mode, respectively. From the dispersion relation $\omega=\sqrt{m^2+|\mathbf{k}|^2}$, the magnitude of group velocity $\mathbf{v}_g\equiv\nabla_\mathbf{k}\omega=\mathbf{k}/\omega$ is given by $v_g^{(n)}=k_n/\Omega_n=\sqrt{1-(m/\Omega_n)^2}$. The energy and the linear momentum and the angular momentum are the conserved charges associated with the translational and rotational symmetries of the action in flat spacetime. From the energy-momentum tensor (EMT), we obtain the energy density $\rho=T_{00}$ and the linear momentum density $p_i=T^{0i}=-T_{0i}$. Note that $T^{0i}$ is also the instantaneous energy flux density; for a $\mathbf{k}$-mode, it equals $\rho \mathbf{v}_g$ under time averaging. The angular momentum density $j_i$ of the scalar and vector fields is derived in Appendix~\ref{App1}.

In the following, we first derive explicit expressions for the radiation fluxes---Eqs.~\eqref{scalar_tau}, \eqref{scalar_P}, and \eqref{scalar_F} in the scalar case and Eqs.~\eqref{vector_tau}, \eqref{vector_P}, and \eqref{vector_F} in the vector case---then consider the dipole approximation and finally specialize to a charged Keplerian binary as the source.

\subsection{Scalar field}
For the scalar field
\begin{align}
\rho&=\frac{1}{2}\left[\dot{\phi}^2+(\partial_i\phi)(\partial_i\phi)+m_\phi^2\phi^2\right]
;
\\
p_i & =-\dot\phi\,\partial_i\phi
;
\\
j_i&=\epsilon_{ikl} x^k \left(-\dot \phi\partial_l\phi\right).
\end{align}
Defining for convenience $D_l\equiv \partial_l + ik_n n^l$ and $D_l^*\equiv\partial_l - ik_n n^l$, the time-averaged angular momentum density of the radiation field \eqref{scalar_radiation_field} reads
\begin{equation}
\begin{aligned}
\frac{r^2}{T}\int_0^T dt\, j_i &=
\frac{r^2}{T}\int_0^T dt\,\epsilon_{ikl} x^k \left(-\dot \phi\partial_l\phi\right)
\\
&=\frac{r}{16\pi^2}\int_0^T \frac{dt}{T}\,\epsilon_{ikl} n^k\sum_{n,s}e^{i\left[\mathbf{k}^{(n)}\cdot\mathbf{x}-\Omega_nt\right]}
e^{-i\left[\mathbf{k}^{(s)}\cdot\mathbf{x}-\Omega_st\right]}i\Omega_n n_{(n)}  D_l^* n_{(s)}^*
\\
&=
\frac{r}{16\pi^2}\epsilon_{ikl} n^k\sum_{n} i\Omega_n  n_{(n)}\partial_l n_{(n)}^*
.
\end{aligned}
\end{equation}
Hence, we obtain
\begin{equation}
\begin{aligned}
\tau_i & =  \int d\Omega_\mathbf{n}
\frac{r}{16\pi^2}\epsilon_{ikl} n^k\sum_{n} i k_n  n_{(n)}\partial_l n_{(n)}^*
.\label{scalar_tau}
\end{aligned}
\end{equation}

The time-averaged energy density is given by
\begin{equation}
\begin{aligned}
\frac{r^2}{T}\int_0^T dt\,\rho &= 
\frac{r^2}{T}\int_0^T dt\,
\frac{1}{2}\left[\dot{\phi}^2+(\partial_l\phi)(\partial_l\phi)+m_\phi^2\phi^2\right]
\\
&=\frac{1}{16\pi^2}\sum_{n}\frac{1}{2} \left[\left(\Omega_n^2+m_\phi^2\right)\left| n_{(n)}\right|^2+ |D_l n_{(n)}|^2 \right]
\\
& \approx 
\frac{1}{16\pi^2}\sum_{n}\Omega_n^2\left| n_{(n)}\right|^2
;
\end{aligned}
\end{equation}
where we used $D_l n_{(n)}\approx ik_n n^l n_{(n)}$. Thus,
\begin{equation}
\begin{aligned}
P & =
\int d\Omega_\mathbf{n} \frac{1}{16\pi^2}\sum_{|n|\ge n_0}\Omega_n k_n \left| n_{(n)}\right|^2
\\
&=
\frac{1}{8\pi^2}\sum_{n\ge n_0}\Omega_n^2\left(1-\frac{n_0^2}{n^2}\right)^{1/2}\int d\Omega_\mathbf{n}  \left| n_{(n)}\right|^2\label{scalar_P}
.
\end{aligned}
\end{equation}
This agrees with the result derived from the instantaneous energy flux density~\cite{PhysRevD.49.6892} as well as the momentum-space calculation~\cite{Mohanty_1996,Cheng:2023qys}. The time-averaged linear momentum density is given by
\begin{equation}
\begin{aligned}
\frac{r^2}{T}\int_0^T dt\,p_i &= 
\frac{1}{16\pi^2}\sum_{n} i\Omega_n  n_{(n)}D_i^* n_{(n)}^*
\\
&\approx 
\frac{1}{16\pi^2}\sum_{n} k_n\Omega_n  |n_{(n)}|^2 n^i
.
\end{aligned}
\end{equation}
Hence, we obtain
\begin{equation}
\begin{aligned}
\mathcal{F}_i & =
\int d\Omega_\mathbf{n} \frac{1}{16\pi^2}\sum_{|n|\ge n_0} k_n^2  |n_{(n)}|^2 n^i
\\
&= 
\frac{1}{8\pi^2}\sum_{n\ge n_0}\Omega_n^2\left(1-\frac{n_0^2}{n^2}\right) \int d\Omega_\mathbf{n}\, |n_{(n)}|^2 n^i
.
\label{scalar_F}
\end{aligned}
\end{equation}

\subsection{Vector field}
For the vector field
\begin{align}
\rho&=\frac{1}{2} \left[\dot A_i \dot A_i+(\partial_i A_0)(\partial_i A_0)+(\partial_i A_j)(\partial_i A_j)-(\partial_i A_j)(\partial_j A_i)\right] +\frac{1}{2}m_A^2 (A_0^2+A_iA_i)
;
\\
p_i & = -(\dot A_j-\partial_j A_0)(\partial_i A_j-\partial_j A_i)-m^2_A A_0A_i
;
\\
j_i&=\epsilon_{ikl}x^{k}\left(-\dot A_{j}\partial_{l}A_{j}+\partial_j A_0\,\partial_l A_j\right)+\epsilon_{ikl}A_{k}\left(\dot A_{l}-\partial_l A_0\right)
.
\end{align}
The time-averaged angular momentum density of the radiation field \eqref{vector_radiation_field} reads
\begin{equation}
\begin{aligned}
\frac{r^2}{T}\int_0^T dt\, j_i&=
\frac{r^2}{T}\int_0^T dt\,
\left[
\epsilon_{ikl}x^{k}\left(-\dot A_{j}\partial_{l}A_{j}+\partial_j A_0\,\partial_l A_j\right)+\epsilon_{ikl}A_{k}\left(\dot A_{l}-\partial_l A_0\right)
\right]
\\
&=\sum_n \frac{1}{16\pi^2}\epsilon_{ikl}\left\{
rn^k\left[i\Omega_n J^{j*}_{(n)}D_{l}J^{j}_{(n)}-D_j J^0_{(n)}\,D_l^* J^{j*}_{(n)}\right]
+
J^k_{(n)}\left[i\Omega_n J^{l*}_{(n)}+D_l^* J^{0*}_{(n)}\right]
\right\}
\\
&\approx
\sum_n \frac{1}{16\pi^2}\epsilon_{ikl}\left\{
rn^k i\Omega_n J^{j*}_{(n)}\partial_l J^{j}_{(n)}
+
J^k_{(n)}\left[i\Omega_n J^{l*}_{(n)}-ik_n n^l J^{0*}_{(n)}\right]
\right\}
\\
&=
\sum_n \frac{1}{16\pi^2} i \Omega_n \epsilon_{ikl}\left\{
rn^k J^{j*}_{(n)}\partial_l J^{j}_{(n)}
+
J^k_{(n)}\left[J^{l*}_{(n)}-\frac{k_n^2}{\Omega_n^2} n^ln^j J^{j*}_{(n)}\right]
\right\}
.
\end{aligned}
\end{equation}
Hence, we obtain
\begin{equation}
\begin{aligned}
\tau_i & =
\int d\Omega_\mathbf{n} \sum_{n\ge n_0} \frac{1}{8\pi^2} i k_n \epsilon_{ikl}\left\{
rn^k J^{j*}_{(n)}\partial_l J^{j}_{(n)}
+
J^k_{(n)}\left[J^{l*}_{(n)}-\left(1-\frac{n_0^2}{n^2}\right) n^ln^j J^{j*}_{(n)}\right]
\right\}
.\label{vector_tau}
\end{aligned}
\end{equation}
The time-averaged energy density is given by
\begin{equation}
\begin{aligned}
\frac{r^2}{T}\int_0^T dt\,\rho &=
\frac{r^2}{T}\int_0^T dt\,
\left\{\frac{1}{2} \left[\dot A_i \dot A_i+(\partial_i A_0)(\partial_i A_0)+(\partial_i A_j)(\partial_i A_j)-2\dot A_i(\partial_i A_0)-(\partial_i A_j)(\partial_j A_i)\right] +\frac{1}{2}m_A^2 (A_0^2+A_iA_i)\right\}
\\
&=\frac{1}{16\pi^2}\sum_n \frac{1}{2} \left\{
(\Omega_n^2+m_A^2)J_{(n)}^{i*}J_{(n)}^i
+\left[D_i J^0_{(n)}\right]\left[D_i J^0_{(n)}\right]^*
+\left[D_i J^j_{(n)}\right]\left[D_i J^j_{(n)}\right]^*
\right.
\\
&\left.\qquad\qquad\qquad\quad +2 (-i\Omega_n)J^i_{(n)}\left[D_i J^0_{(n)}\right]^*
-\left[D_i J^j_{(n)}\right]\left[D_j J^i_{(n)}\right]^*
\right\}
+\frac{1}{2}m_A^2 |J^0_{(n)}|^2 
\\
& \approx \frac{1}{16\pi^2}\sum_n \left[
\Omega_n^2 J_{(n)}^{i*}J_{(n)}^i
-k_n^2 n^i n^j J^{i*}_{(n)}J^j_{(n)}
\right]
.
\end{aligned}
\end{equation}
Hence, we obtain
\begin{equation}
\begin{aligned}
P & =
\int d\Omega_\mathbf{n} \frac{1}{16\pi^2}\sum_{|n|\ge n_0} \left[
\Omega_n^2 J_{(n)}^{i*}J_{(n)}^i
-k_n^2 n^i n^j J^{i*}_{(n)}J^j_{(n)}
\right](k_n/\Omega_n)
\\
& =
\frac{1}{8\pi^2}\sum_{n\ge n_0} \Omega_n^2 \left(1-\frac{n_0^2}{n^2}\right)^{1/2}
\int d\Omega_\mathbf{n}\left[
J_{(n)}^{i*}J_{(n)}^i
-\left(1-\frac{n_0^2}{n^2}\right) n^i n^j J^{i*}_{(n)}J^j_{(n)}
\right]
.\label{vector_P}
\end{aligned}
\end{equation}
This agrees with the result derived from the instantaneous energy flux density~\cite{PhysRevD.49.6892} as well as the momentum-space calculation~\cite{KumarPoddar:2019ceq,Cheng:2023qys}. The time-averaged linear momentum density is given by
\begin{equation}
\begin{aligned}
\frac{r^2}{T}\int_0^T dt\,p_i &= 
\frac{r^2}{T}\int_0^T dt\,
\left[
 (\dot A^j+\partial_j A^0)(\partial_j A^i-\partial_i A^j)+m^2_A A^0A^i
\right]
\\
&=
\frac{1}{16\pi^2}
\sum_n \left\{
\left[ i\Omega_n J_{(n)}^{j*} +D_j^*J_{(n)}^{0*}\right]
\left[ D_j J_{(n)}^i -D_iJ_{(n)}^j\right]
+m_A^2 J_{(n)}^{0*} J_{(n)}^{i}
\right\}
\\
&\approx
\frac{1}{16\pi^2}
\sum_n \left\{
\left[ i\Omega_n J_{(n)}^{j*}-ik_n n^jJ_{(n)}^{0*}\right]
ik_n\left[ n^j J_{(n)}^i -n^iJ_{(n)}^j\right]
+m_A^2 J_{(n)}^{0*} J_{(n)}^{i}
\right\}
\\
&=
\frac{1}{16\pi^2}
\sum_n \left\{
k_n\Omega_n\left[\frac{k_n^2}{\Omega_n^2} n^j n^l J_{(n)}^{l*}- J_{(n)}^{j*}\right]
\left[ n^j J_{(n)}^i -n^iJ_{(n)}^j\right]
+m_A^2 \frac{k_n}{\Omega_n}n^l J_{(n)}^{l*} J_{(n)}^{i}
\right\}
;
\end{aligned}
\end{equation}
hence we obtain
\begin{equation}
\begin{aligned}
\mathcal{F}_i & =
\int d\Omega_\mathbf{n}\, 
\frac{1}{8\pi^2}
\sum_{n\ge n_0} \left\{
k_n^2\left[\frac{k_n^2}{\Omega_n^2} n^j n^l J_{(n)}^{l*}- J_{(n)}^{j*}\right]
\left[ n^j J_{(n)}^i -n^iJ_{(n)}^j\right]
+m_A^2 \frac{k_n^2}{\Omega_n^2}n^l J_{(n)}^{l*} J_{(n)}^{i}
\right\}
.
\label{vector_F}
\end{aligned}
\end{equation}

Note that in the massless limit, Eq.~\eqref{vector_radiation_field} gives the radiation field in the Lorenz gauge (see also \cite{Cardoso_2016}). In this case, the computation can be equivalently performed in the radiation gauge ($A_0=\partial_iA_i=0$)~\cite{GW_volume_1,Liu_2020,Christiansen_2021}.

\subsection{Dipole radiation}
For $|\mathbf{k}^{(n)}\cdot\mathbf{x}|\ll 1$, Eq.~\eqref{J^a_n} can be approximated by
\begin{align}
J^i_{(n)} &\approx \int d^3x\left[\frac{1}{T}\int_0^Tdt\,J^i(t,\mathbf{x})\,e^{i\Omega_n t}\right]\equiv \Omega_n \,j_n^i,
\\
J^0_{(n)} &\approx \int d^3x\,\left[-i\mathbf{k}^{(n)}\cdot\mathbf{x}\right]\left[\frac{1}{T}\int_0^Tdt\,J^0(t,\mathbf{x})\,e^{i\Omega_n t}\right]
= \mathbf{j}_n\cdot \mathbf{k}^{(n)},
\end{align}
which gives rise to the electric dipole radiation. The next-to-leading order term in the expansion of $e^{-i\mathbf{k}\cdot\mathbf{x}}$ gives rise to, e.g., the electric quadrupole radiation~\cite{PhysRevD.49.6892,Cheng:2023qys}. This is subleading when the dipole moment is sufficiently large and will be neglected in this paper. As mentioned earlier, we also neglect the PN corrections to the source term~\cite{Huang:2018pbu,Bhattacharyya:2023kbh,Diedrichs:2023foj,Henry_2024}. Similarly, the dipole approximation to the scalar source \eqref{n_{(n)}} is
\begin{align}
n_{(n)} &\approx \int d^3x\,\left[-i\mathbf{k}^{(n)}\cdot\mathbf{x}\right]\left[\frac{1}{T}\int_0^Tdt\,n(t,\mathbf{x})\,e^{i\Omega_n t}\right]= \mathbf{j}_n\cdot \mathbf{k}^{(n)}.
\end{align}

Using $\partial_l n^j=(\delta_{lj}-n^ln^j)/r$, and
\begin{equation}
	\begin{aligned}
        \int d\Omega_\mathbf{n}\, n_i n_j =\frac{4\pi}{3}\delta_{ij},
        \quad
		\int d\Omega_\mathbf{n}\, n_i n_j n_k n_l =\frac{4\pi}{15}(\delta_{ij}\delta_{kl}+\delta_{ik}\delta_{jl}+\delta_{il}\delta_{jk}),
	\end{aligned}
\end{equation}
Equation~\eqref{vector_tau} gives the angular momentum flux of the vector dipole radiation:
\begin{equation}
\begin{aligned}
\tau_i
&=
\sum_{n\ge n_0} \frac{i}{8\pi^2}\,k_n\Omega_n^2\,\epsilon_{ikl} \int d\Omega_\mathbf{n}\,
j^k_n\left[j^{l*}_{n}-\left(1-\frac{n_0^2}{n^2}\right) n^ln^j\, j^{j*}_{n}\right]
\\
&=
\frac{i}{6\pi}\sum_{n\ge n_0} \Omega_n^3\left(1-\frac{n_0^2}{n^2}\right)^{1/2}
\left(2+\frac{n_0^2}{n^2}\right) \epsilon_{ikl}\,j^k_nj^{l*}_{n} \label{vector_dip_tau}
.
\end{aligned}
\end{equation}
The associated energy flux from Eq.~\eqref{vector_P} is
\begin{equation}
\begin{aligned}
P &=\frac{1}{6\pi}\sum_{n\ge n_0} \Omega_n^4 \left(1-\frac{n_0^2}{n^2}\right)^{1/2}
\left(2+\frac{n_0^2}{n^2}\right)j^{i*}_n\,j^i_n. \label{vector_dip_P}
\end{aligned}
\end{equation}
Equation~\eqref{scalar_tau} gives the angular momentum flux of the scalar dipole radiation:
\begin{equation}
\begin{aligned}
\tau_i & =  \int d\Omega_\mathbf{n}
\frac{r}{8\pi^2}\epsilon_{ikl}\,n^k\sum_{n\ge n_0} i\,k_n^3\, n^j\,j^j_n\,(\partial_l n^s)\, j_n^{s*}
\\
&= \int d\Omega_\mathbf{n}
\frac{1}{8\pi^2}\epsilon_{ikl} \sum_{n\ge n_0} i\,k_n^3\,(n^kn^j\delta_{ls}-n^kn^jn^ln^s)\, j^j_n\,j_n^{s*}
\\
&=
\frac{i}{6\pi}\sum_{n\ge n_0} \Omega_n^3\left(1-\frac{n_0^2}{n^2}\right)^{3/2} \epsilon_{ikl}\,j^k_n\,j^{l*}_n \label{scalar_dip_tau}
.
\end{aligned}
\end{equation}
The associated energy flux from Eq.~\eqref{scalar_P} is
\begin{equation}
\begin{aligned}
P
=\frac{1}{6\pi}\sum_{n\ge n_0}\Omega_n^4\left(1-\frac{n_0^2}{n^2}\right)^{3/2}  j^{i*}_n\,j^i_n \label{scalar_dip_P}
.
\end{aligned}
\end{equation}

As can be seen from Eqs.~\eqref{scalar_F} and \eqref{vector_F}, the linear momentum flux of the dipole radiation vanishes.

\subsection{Charged binary}\label{sec:Charged binary}
We now apply these results to a charged Keplerian binary. We model the binary as two point charges. This description is valid so long as the size $R$ of the charge distribution satisfies
\begin{equation}
mR=0.1\left(\frac{m}{1.3\times 10^{-21}\,\text{eV}}\right)\left(\frac{R}{10^{10}M_\odot}\right)\ll 1.
\end{equation}
At the Newtonian order, the dark charges induce a Yukawa force between the two bodies. The interaction potential is given by $-\frac{\pm q_1q_2}{4\pi r}e^{-mr}$, with $+$ ($-$) corresponding to the scalar (vector) case. For $mr \ll 1$, the equation of motion of $\mathbf{X}(t)=(X,Y,Z)$ can be approximated by
\begin{equation}\label{EOM}
\begin{aligned}
\ddot{\mathbf{X}} &=-\nabla\left[-\frac{M}{r}\left(1+\alpha_\phi e^{-m_\phi r}-\alpha_A e^{-m_A r}\right)\right]
\\
&= -\frac{M}{r^3}\left[1+\alpha_\phi (1+m_\phi r)e^{-m_\phi r}-\alpha_A (1+m_A r)e^{-m_A r}\right]\mathbf{X}
\\
&= -\frac{M}{r^3}\left[1+\alpha_\phi\left(1-\frac{1}{2}m_\phi^2 r^2\right)-\alpha_A\left(1-\frac{1}{2}m_A^2 r^2\right) + \mathcal{O}( m_\phi^3 r^3)+ \mathcal{O}( m_A^3 r^3)\right]\mathbf{X}
\\
&\approx -\frac{\tilde M}{r^3}\mathbf{X},
\end{aligned}
\end{equation}
with $M\equiv m_1+m_2$, $\alpha \equiv q_1q_2/(4\pi m_1m_2)$, and $\tilde M\equiv M(1+\alpha_\phi-\alpha_A)$. The binary orbit in this approximation is thus Keplerian. If $\frac{1}{2}(mr)^2\ll 1$ is not satisfied, the Keplerian approximation can also be applied if $\alpha$ is sufficiently small such that the Yukawa force is negligible, similar to the PN corrections in the pure gravitational sector. Otherwise, we have to consider the exact 0PN orbit (aperiodic in the noncircular case), for which the dark radiation is suppressed, since $n_0\gg m r$ and the orbital evolution is mainly driven by the gravitational radiation.

We consider an elliptical orbit with semimajor axis $a$ and eccentricity $e$, which can be parametrized by the eccentric anomaly $\xi$ as
\begin{equation}
X(t)=a(\cos\xi-e),\quad Y(t)=a\sqrt{1-e^2}\sin\xi,\quad Z(t)=0,\quad\Omega t=\xi-e\sin\xi;
\end{equation}
with $t=0$ the periastron-crossing time and the orbital angular momentum along the $Z$-direction. The mean orbital frequency is given by $\Omega=2\pi/T=\sqrt{{\tilde M}/{a^3}}$. Note that
\begin{equation}
\frac{1}{2}(m a)^2 \approx 0.01\left(\frac{m}{10^{-24}\,\text{eV}}\right)^2\left(\frac{a}{1\,\text{pc}}\right)^2,
\quad
\frac{\Omega}{2\pi}\approx 0.03\left(\frac{\tilde M}{10^{10}M_\odot}\right)^{1/2}\left(\frac{a}{1\,\text{pc}}\right)^{-3/2}\text{nHz}.
\end{equation}

In the center-of-mass (CM) frame of the binary, $\mathbf{X}_1=({m_2}/{M})\mathbf{X}$ and $\mathbf{X}_2=-({m_1}/{M})\mathbf{X}$. Using the integral representation of the Bessel function of the first kind
\begin{equation}
J_n(z)=\frac1{2\pi}\int_0^{2\pi} d\xi\,e^{i\left(n\xi-z\sin\xi\right)},
\end{equation}
the source term of the dipole radiation can be evaluated in the CM frame as
\begin{equation}
\mathbf{j}_n=
\mu \gamma a\,
\frac{1}{n}\left(\begin{matrix}
-i J'_n
\\
\frac{\sqrt{1-e^2}}{e}J_n
\\
0
\end{matrix}\right)
,
\end{equation}
where $J_n\equiv J_n(ne)$, $J_n'\equiv \left[{dJ_n(z)}/{dz}\right]_{z=ne}=(J_{n-1}-J_{n+1})/2$, $\mu\equiv m_1m_2/M$, and $\gamma\equiv {q_1}/{m_1}-{q_2}/{m_2}$. The dipole approximation is valid because the radiation wavelength $\lambda_n=2\pi/k_n\ge 2\pi/n\Omega\sim\sqrt{a^3/\tilde M}\gg a$.

From Eqs.~\eqref{vector_dip_tau}--\eqref{scalar_dip_P}, we obtain the energy flux $ P_\text{dip}$ and the angular momentum flux $\boldsymbol{\tau}_\text{dip}$ of the dipole radiation,
\begin{align}
 P_\text{dip} &=\frac{1}{6\pi}(\gamma\mu a)^2\Omega^4 \sum_{n\ge n_0} n^2\left[(J_n')^2+\frac{1-e^2}{e^2}(J_n)^2\right]\Upsilon_n
 ;\label{P_dip}
 \\
 \boldsymbol{\tau}_\text{dip}&=\frac{1}{6\pi}(\gamma\mu a)^2\Omega^3\sum_{n\ge n_0}2n\left(\sqrt{\frac{1-e^2}{e^2}}J_n'J_n\right)\Upsilon_n\,\mathbf{e}_Z
 ;\label{tau_dip}
\end{align}
where $n_0=m/\Omega$, and
\begin{equation}\label{eq:scalar_vector}
\Upsilon_n=\begin{cases}\left(1-\frac{n_0^2}{n^2}\right)^{3/2},&\text{(scalar)}
\\
\left(1-\frac{n_0^2}{n^2}\right)^{1/2}\left(2+\frac{n_0^2}{n^2}\right)\,\,\,,&\text{(vector)}
\end{cases}
\end{equation}
For the circular orbit ($e=0$), $\lim_{e\to 0} J_n'=\lim_{e\to 0} \sqrt{\frac{1-e^2}{e^2}}J_n=\frac{1}{2}\delta_{n,1}$, hence $P_\text{dip}=\tau_\text{dip}\Omega=\frac{1}{12\pi}(\gamma\mu a)^2\Omega^4\,\Upsilon_1$. The enhancement of $P_\text{dip}$ for $n_0<1$ due to eccentricity is depicted in Fig.~\ref{fig:P}. In the case of massless radiation ($n_0=0$), using
\begin{align}
\sum_{n=1}^\infty n^2(J_n')^2 =\frac{1+{3 e^2}/{4}}{4 \left(1-e^2\right)^{5/2}}
,
\quad
\sum_{n=1}^\infty n^2(J_n)^2 = \frac{e^2 \left(1+{e^2}/{4}\right)}{4 \left(1-e^2\right)^{7/2}}
,
\quad
\sum_{n=1}^\infty nJ_n' J_n =\frac{e}{4 \left(1-e^2\right)^{3/2}}
,
\end{align}
we obtain
\begin{align}
P_\text{dip} =\frac{1}{12\pi}(\gamma\mu a)^2\Omega^4 \frac{1+e^2/2}{(1-e^2)^{5/2}}\Upsilon
,
\quad
\tau_\text{dip} =\frac{1}{12\pi}(\gamma\mu a)^2\Omega^3 \frac{1}{1-e^2}\Upsilon
, \label{dip_massless}
\end{align}
with $\Upsilon=1$ for the scalar field and $\Upsilon=2$ for the vector field, recovering the correct massless limits~\cite{Liu_2020,Christiansen_2021,Cardoso_2021}.

Meanwhile, the leading-order energy and angular momentum fluxes of gravitational radiation are given by~\cite{PhysRev.131.435,PhysRev.136.B1224}
\begin{align}
P_\text{gw}
&=\frac{32}{5}\mu^2 a^4\Omega^6 \sum_{n=1}^\infty g(n,e)
=\frac{32}{5}\mu^2a^4 \Omega^6\frac{1+\frac{73}{24}e^2+\frac{37}{96}e^4}{(1-e^2)^{7/2}},
\\
\tau_\text{gw}
&=
\frac{32}{5}\mu^2 a^4\Omega^5 \sum_{n=1}^\infty h(n,e)
=
\frac{32}{5}\mu^2 a^4\Omega^5\frac{1+\frac{7}{8}e^2}{(1-e^2)^2}
,
\end{align}
where
\begin{align}
g(n,e)&=\frac{n^2}{2}\left\{J_n^2\left[n^2\frac{(1-e^2)^3}{e^4}+\frac{3+e^4-3e^2}{3e^4}\right]+(J_n')^2\left[n^2\frac{(1-e^2)^2}{e^2}+\frac{1-e^2}{e^2}\right]+n J_n J_n'\frac{-4+7e^2-3e^4}{e^3}\right\}
,\label{g}
\\
h(n,e) &=
n^2 J_n^2\frac{\left(e^2-2\right) \left(1-e^2\right)^{3/2}}{e^4}
-n^2 (J_n')^2\frac{2 \left(1-e^2\right)^{3/2}}{e^2}
+nJ_nJ_n'\left[n^2\frac{2 \left(1-e^2\right)^{5/2}}{e^3}+\frac{\sqrt{1-e^2} \left(2-e^2\right)}{e^3}\right]
.
\end{align}

\section{Adiabatic orbital evolution}
\label{sec.3}
The binary orbit slowly evolves due to radiation reaction. In the present approximation, the short-term orbital motion of the binary is Keplerian, with the orbital energy $E$ and angular momentum $\mathbf{L}$ at Newtonian order given by
\begin{equation}
E=\mu\left(\frac{1}{2}v^2-\frac{\tilde M}{r}\right)=-\frac{\mu \tilde M}{2a},
\quad
\mathbf{L}=\mu\,\mathbf{X}\times\mathbf{v}=\mu\sqrt{(1-e^2)a \tilde M}\,\mathbf{e}_Z.
\end{equation}
The flux-balance equations relating the radiation flux integrated over one orbital period to the changes in the energy and angular momentum of the source read\footnote{A first-principle treatment would require explicitly calculating the radiation-reaction force, which is beyond the scope of this paper. The relative acceleration in the CM frame due to gravitational radiation reaction at leading 2.5PN order, consistent with $P_\text{gw}$, $\tau_\text{gw}$ and the result for an uncharged binary in the harmonic gauge, is $\ddot{\mathbf{X}}\supset\frac{\tilde M}{r^3}\frac{24\mu}{5}\left[\left(v^{2}+\frac{17\tilde M}{9r}\right)\dot{r}\,\frac{\mathbf{X}}{r} -\left(\frac{v^{2}}{3}+\frac{\tilde M}{r}\right)\mathbf{v}\right]$.}
$P+\dot E=\tau_i+\dot L_i=0$. Consequently, the secular rates of change of $\{a,e,\Omega\}$ in the adiabatic approximation are
\begin{align}\label{orbital_evolution}
\dot a =-\frac{2a^2P}{\mu \tilde M}
;
\quad
\dot e =-\left(\frac{1-e^2}{e}\right)\frac{a}{\mu \tilde M}\left(P-\frac{\tau \Omega}{\sqrt{1-e^2}}\right)
;
\quad
\dot \Omega = -\frac{3}{2}\frac{\tilde M^{1/2}}{a^{5/2}}\dot a.
\end{align}
In the present case, $P=P_\text{gw}+P_\text{dip}$, and $\tau=\tau_\text{gw}+\tau_\text{dip}$, with $P_\text{dip}$ and $\tau_\text{dip}=|\boldsymbol{\tau}_\text{dip}|$ given by Eqs.~\eqref{P_dip}--\eqref{tau_dip}. For $m=0$, Eq.~\eqref{dip_massless} leads to $P_\text{dip}-\tau_\text{dip}\Omega/\sqrt{1-e^2}=\frac{1}{12\pi}(\gamma\mu a)^2\Omega^4 \frac{3 e^2}{2 \left(1-e^2\right)^{5/2}}\Upsilon$; for $m>0$, the contribution of the $n$th harmonic mode to $P_\text{dip}-\tau_\text{dip}\Omega/\sqrt{1-e^2}$ is negative only when $n$ is sufficiently small or the eccentricity is sufficiently large. However, this contribution is subdominant, and hence, $P_\text{dip}-\tau_\text{dip}\Omega/\sqrt{1-e^2}>0$. The scalar or vector dipole radiation thus tends to circularize the orbit, as does the gravitational radiation. Compared with the 0PN gravitational radiation, the dipole radiation enters formally at $-1$PN order for $\Omega\gg m$, and is suppressed by the boson mass $m$. The coupling strength $\alpha$ affects the orbital evolution through the modified Keplerian relation $a=(\tilde M/\Omega^2)^{1/3}$, with $\tilde M=M(1\pm \alpha)$. At a given $\{\Omega,e\}$,
\begin{equation}
\frac{\mathrm{d}e}{\mathrm{d}a}\propto \frac{1}{\tilde M^{1/3}}\left(1-\frac{\tau\Omega/P}{\sqrt{1-e^2}}\right),
\quad
\frac{\mathrm{d}e}{\mathrm{d}\Omega}\propto 1-\frac{\tau\Omega/P}{\sqrt{1-e^2}},
\quad
\frac{\tau}{P}=\frac{\tau_\text{gw}}{P_\text{gw}}\left(1+\frac{
{P_\text{gw}}/{\tau_\text{gw}}-{P_\text{dip}}/{\tau_\text{dip}}}{
{P_\text{gw}}/{\tau_\text{dip}}+{P_\text{dip}}/{\tau_\text{dip}}}\right)
.
\end{equation}
Since $P_\text{gw}/\tau_\text{dip}\propto a^2\propto \tilde M^{2/3}$, the ratio $\tau/P$ decreases (increases) in the scalar (vector) case if $\alpha$ increases, leading to a faster (slower) decay of $e$ with respect to $\Omega$. For $Q_1\equiv|q_1/m_1|\ge Q_2\equiv|q_2/m_2|$, the relation between $\{Q_1,Q_2\}$ and $\{\gamma,\alpha\}$ is given by $Q_1=\frac{1}{2}\left(|\gamma|+\sqrt{\gamma^2+16\pi \alpha}\right)$ and $Q_2=\frac{1}{2}\left(-|\gamma|+\sqrt{\gamma^2+16\pi \alpha}\right)\text{sgn}(q_1q_2)$. For concreteness, and focusing on the case of large dipole strength $\gamma$ (where only sufficiently small values of $\alpha$ are admissible if $Q_1<1$), we henceforth fix $\alpha=0.001$ unless otherwise stated.

\begin{figure}[t]
  \centering
  \begin{subfigure}[b]{0.45\textwidth}
    \centering
    \includegraphics[width=\textwidth]{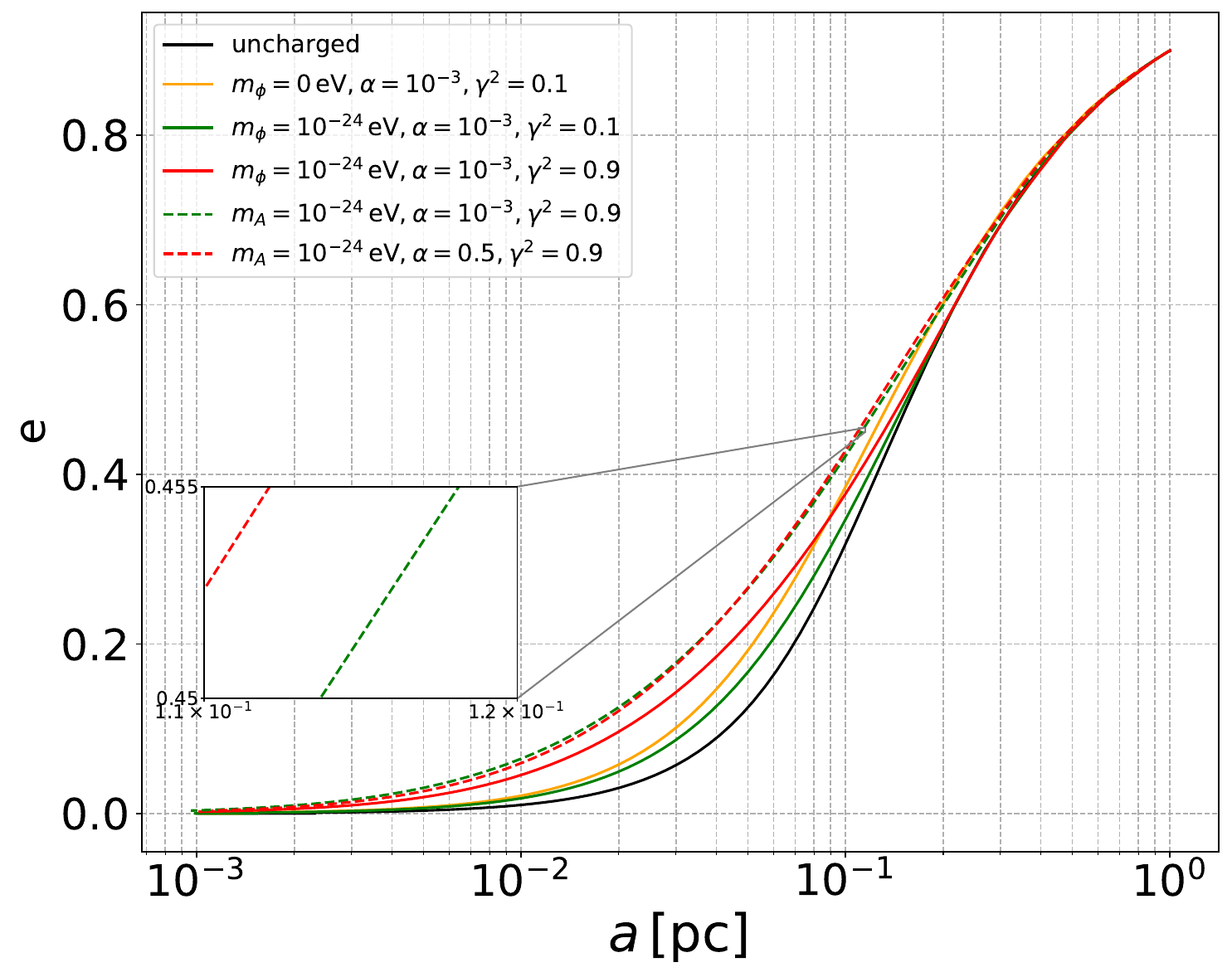}
    \caption{}
    \label{fig:e(a)}
  \end{subfigure}
  \qquad
  \begin{subfigure}[b]{0.45\textwidth}
    \centering
    \includegraphics[width=\textwidth]{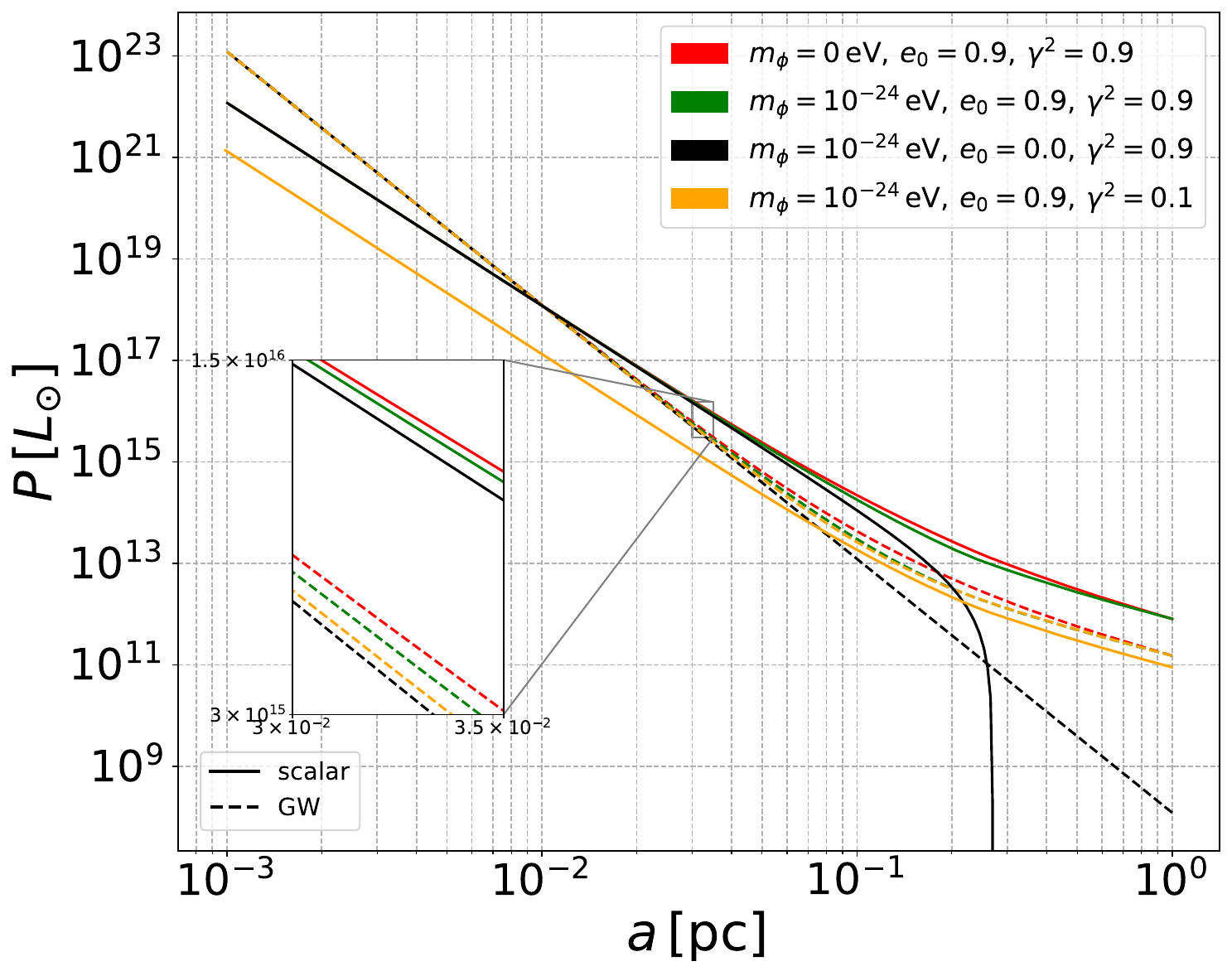}
    \caption{}
    \label{fig:power_compare}
  \end{subfigure}
  \jcaption{
  \textbf{Left:} Eccentricity evolution with semimajor axis starting from $e = 0.9$ and $a = 1\,\mathrm{pc}$. \textbf{Right:} Comparison between $P_\text{gw}$ (dashed lines) and $P_\text{dip}$ (solid lines) during the inspiral in the scalar case, with $\alpha=0$. The energy flux is expressed in units of solar luminosity $L_\odot$. In both panels, we choose $M = 10^{10}\,M_{\odot}$ and $q = 0.1$.
  }
  \label{fig:orbit}
\end{figure}

We set the initial semimajor axis to $1\,\mathrm{pc}$ ($\approx 2\times 10^{13}M_\odot$). Figure~\ref{fig:e(a)} shows the evolution of $\{a,e\}$ under various parameter sets, all starting from an initial eccentricity $e_0 = 0.9$. Solid and dashed lines correspond to the scalar and vector cases, respectively. In the presence of dipole radiation, although $|\dot e|$ is enhanced relative to the uncharged case, the eccentricity initially decays more slowly with respect to $a$ since ${P_\text{gw}}/{\tau_\text{gw}}>{P_\text{dip}}/{\tau_\text{dip}}$ (this holds for all eccentricities in the massless case and for sufficiently high eccentricities in the massive case, as shown in Fig.~\ref{fig:P/tau}). The binary thus becomes more eccentric at a given orbital frequency for the same initial condition. For $\gamma^2\gtrsim 0.1$, the impact of $\alpha$ on the evolution is comparatively small. In contrast to circular orbits, dipole radiation from highly eccentric orbits turns on at $\Omega\ll m$. As the eccentricity increases, the impact of boson mass becomes smaller. Figure~\ref{fig:power_compare} illustrates this behavior for the scalar case, where we compare the energy fluxes of GW and scalar radiation for different values of $e_0$ and $\gamma^2$.

\begin{figure}[t]
  \centering
  \begin{subfigure}[b]{0.52\textwidth}
    \centering
    \includegraphics[width=\textwidth]{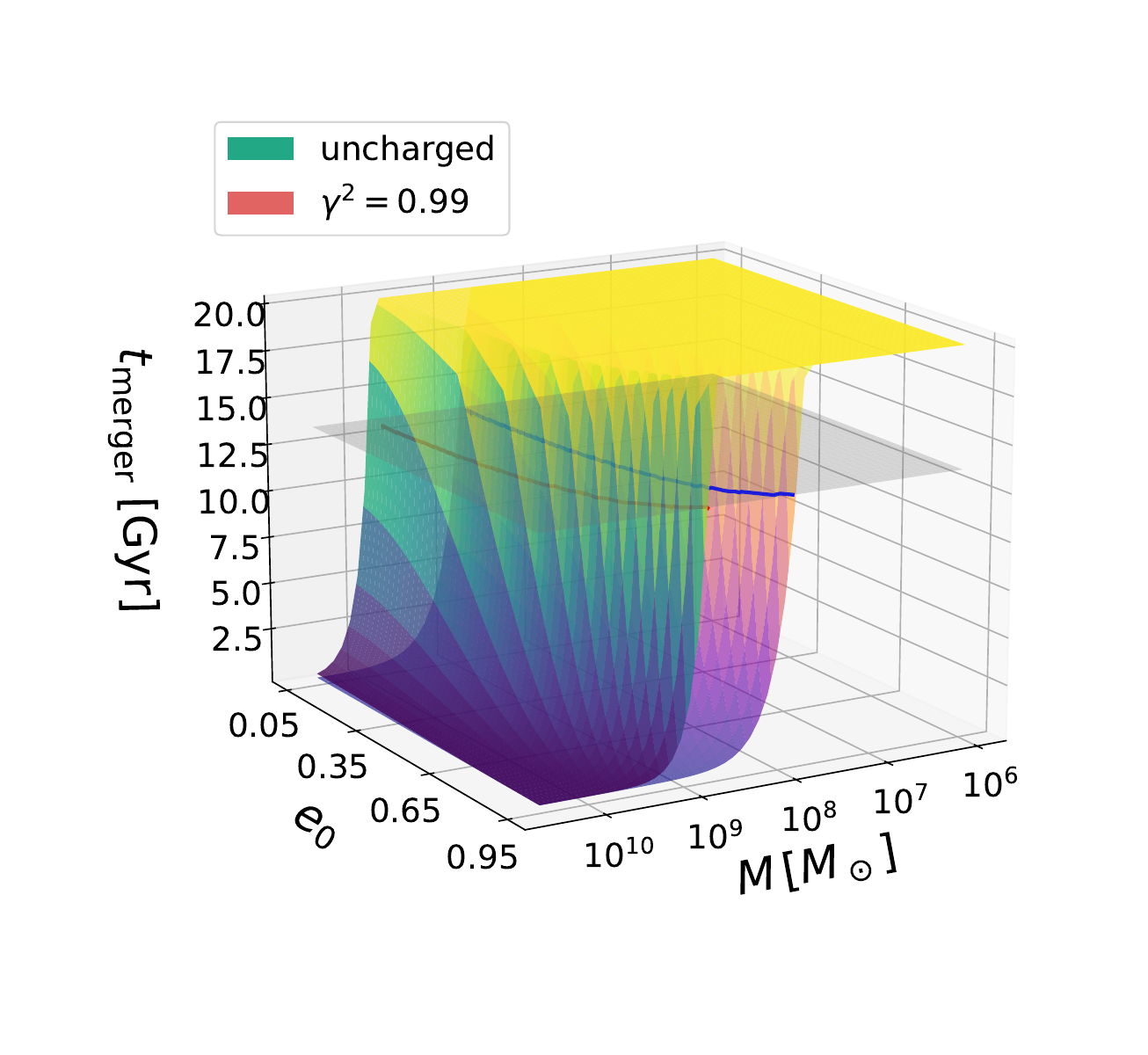}
    \caption{}
    \label{fig:tmerger}
  \end{subfigure}
  \begin{subfigure}[b]{0.47\textwidth}
    \centering
    \includegraphics[width=\textwidth]{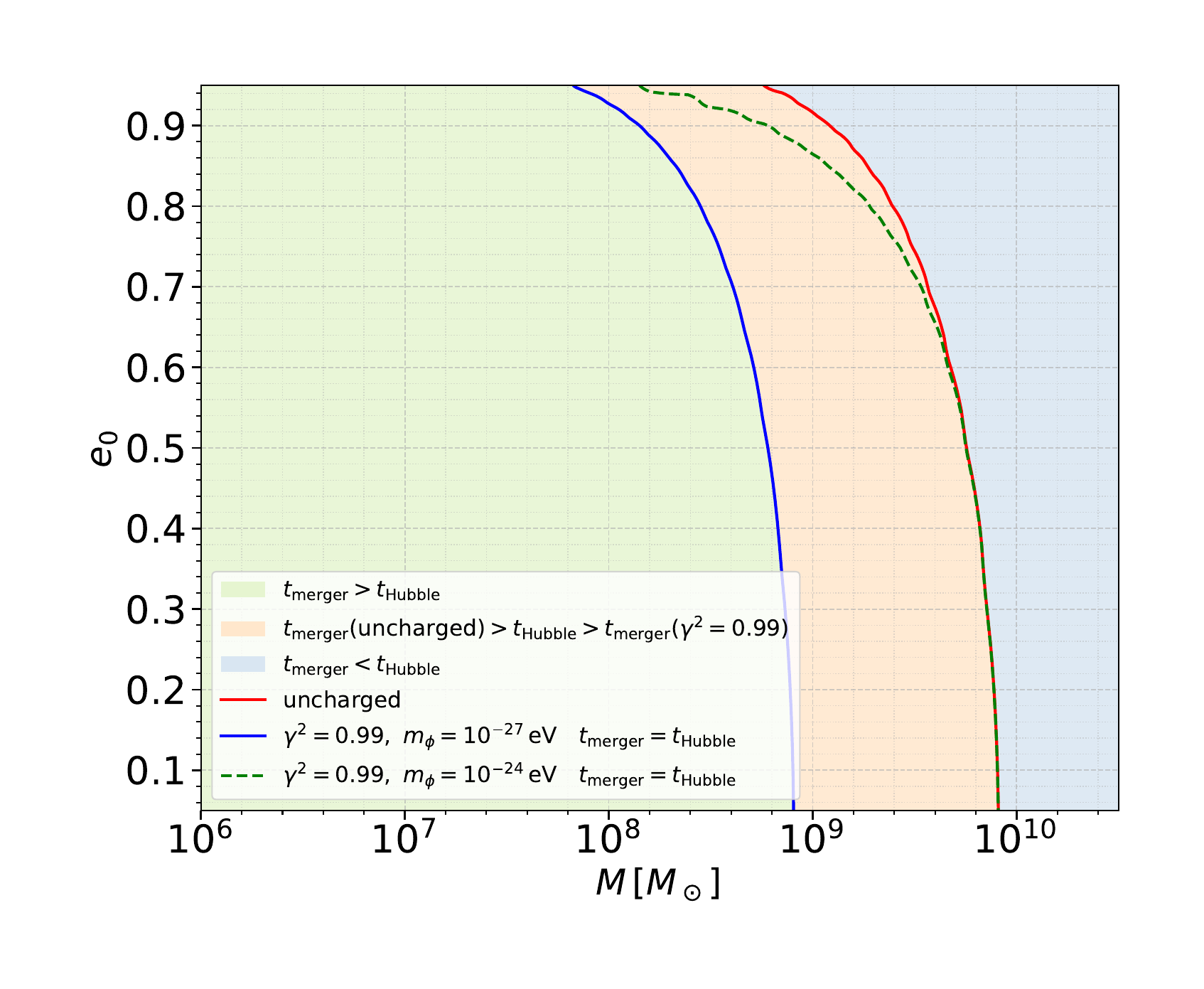}
    \caption{}
    \label{fig:tmerger_2d}
  \end{subfigure}
  \jcaption{\textbf{Left}: SMBHB merger time as a function of $e_0$ and $M$ for $m_\phi=10^{-27}\,\mathrm{eV}$. The gray plane marks the Hubble time, and regions with merger times exceeding $20\,\mathrm{Gyr}$ are truncated. \textbf{Right}: $e_0(M)$ curves corresponding to $t_\text{merger}=t_\text{Hubble}$. The parameter space of successful mergers expands as $\gamma^2$ increases. In both panels, we choose $q=0.1$, and $\alpha=0.001$.}
  \label{fig:time}
\end{figure}

To assess whether dipole radiation can reduce the binary merger time below a Hubble time $t_\text{Hubble} = 1/H_0$ ($H_0$ being the Hubble parameter today) and thereby alleviate the final-parsec problem, we consider systems with total mass $M \in [10^6, 10^{11}]M_{\odot}$ and compute, for various initial eccentricities, the time $t_\text{merger}$ required to evolve from $a=1\,\mathrm{pc}$ to $a=0$. Figure~\ref{fig:tmerger} compares the result in the uncharged case ($\gamma=\alpha=0$) with that for a large dipole strength ($\gamma^2 = 0.99$) in the scalar case. As can be seen, although a large dipole strength can expand the parameter space leading to coalescence within $t_\text{Hubble}$ relative to the uncharged case, it is still insufficient to resolve the final-parsec problem for SMBHBs with sufficiently small total mass $M$, except for binaries with near-maximal eccentricity---an unlikely configuration to be maintained at $a\sim 1\,\mathrm{pc}$ without additional mechanisms. The results for $m_\phi=10^{-27}\,\mathrm{eV}$ are nearly indistinguishable from those in the massless case, whereas for $m_\phi=10^{-24}\,\mathrm{eV}$, even with $\gamma^2=0.99$, the expansion of the parameter space remains quite limited, as shown in Fig.~\ref{fig:tmerger_2d}.

At frequencies as low as a few nanohertz, however, the SGWB is expected to be dominated by binaries with total masses $10^{8}M_\odot\lesssim M \lesssim 10^{10}M_\odot$~\cite{Ellis_2023, gardiner2024backgroundgravitationalwaveanisotropy, PhysRevD.110.063020,Kelley:2017lek}. So long as the charged binaries in this mass range merge within $t_\text{Hubble}$, the nHz-band characteristic strain spectrum can be used to fit our model. For SMBHBs whose merger times exceed $t_\text{Hubble}$, we account for their contribution to the GW background only up to a maximum evolution time of $t_\text{Hubble}$. We find that the resulting reduction in characteristic strain has a minimal effect on the overall spectral shape and amplitude.

\section{Stochastic gravitational wave background and Bayesian analysis}
\label{sec.4}
The SGWB from SMBHB inspirals is the superposition of gravitational waves emitted by the entire population of inspiralling SMBHBs. In the previous sections, we derived the dipole radiation fluxes and examined their effects on the orbital evolution of individual binaries. This modification of the orbital dynamics leaves an imprint on the resulting gravitational wave energy spectrum,
\begin{align}
    \frac{\mathrm{d}E_{\text{gw}}}{\mathrm{d}f_\text{s}} \bigg|_{f_\text{s}
    = (1+z)f} &= \sum_{n=1}^{\infty} \frac{P_\text{gw}^n}{n \dot\Omega/2\pi}\bigg|_{\Omega= (1+z)2\pi f/n}
    \\
    &=\frac{64\pi}{15}\mu^3\tilde M^2\sum_{n=1}^{\infty}\frac{g(n, e)}{n}\frac{\Omega^3}{P}\bigg|_{\Omega= (1+z)2\pi f/n}\label{dE/df_formula}
    ;
\end{align}
where we used Eq.~\eqref{orbital_evolution} and $P_\text{gw}^n = \frac{32}{5}\mu^2 a^4\Omega^6 g(n, e)$, with $g(n,e)$ given by Eq.~\eqref{g}. Note that for given $\{\Omega,e\}$, $\mathrm{d}E_\text{gw}/\mathrm{d}f_\text{s}$ is proportional to $\tilde M^{4/3}$ if the energy flux is dominated by $P_\text{dip}$ and to $\tilde M^{2/3}$ if it is dominated by $P_\text{gw}$. To demonstrate the main effects of dipole radiation, we assume a uniform distribution of the source parameters $\{\gamma^2,e_0,m\}$ ($e_0$ being the eccentricity at $a=1\,\mathrm{pc}$) for simplicity. The characteristic strain spectrum\footnote{The SGWB is assumed to be stationary, isotropic, and unpolarized. The characteristic strain spectrum is defined as $h_c(f)=\sqrt{fS_h(f)}$, with $S_h(f)$ being the strain power spectral density~\cite{romano2023searchesstochasticgravitationalwavebackgrounds}. It corresponds to the normalized energy density spectrum $\Omega_\text{gw}(f)=\frac{1}{\rho_\text{crit}}\frac{d\rho_\text{gw}}{d \ln f}=\frac{2\pi^2}{3H_0^2}f^2h_c^2$, where $\rho_\text{crit}=3H_0^2/(8\pi)$  is the critical density today.} of the SGWB can be expressed by incorporating the merger rate of SMBHBs per comoving volume~\cite{phinney2001practicaltheoremgravitationalwave, Chen_2017},
\begin{equation}
    h^2_c(f;\gamma^2,e_0,m) \approx \frac{4}{\pi f}\int\mathrm{d}z\,\mathrm{d}M\,\mathrm{d}q\,\frac{\mathrm{d}^3n}{\mathrm{d}z\,\mathrm{d}M\,\mathrm{d}q}\,\frac{\mathrm{d}E_{\mathrm{gw}}}{\mathrm{d}f_\text{s}}.
\end{equation}

\begin{figure}[ht]
  \centering
  \begin{subfigure}[b]{0.45\textwidth}
    \includegraphics[width=\textwidth]{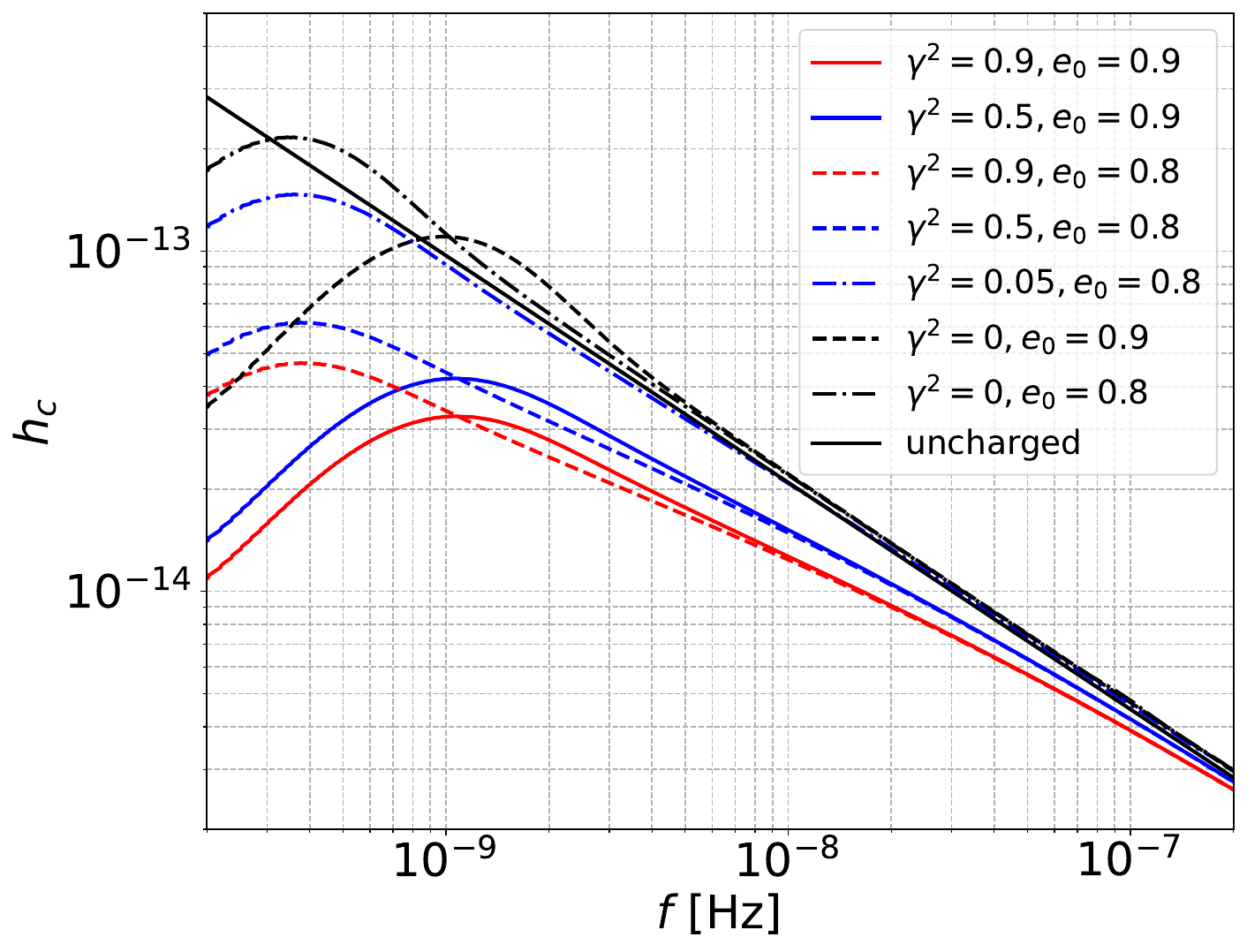}
    \caption{}
    \label{fig:strain_sub1}
  \end{subfigure}%
  \qquad
  \begin{subfigure}[b]{0.45\textwidth}
    \includegraphics[width=\textwidth]{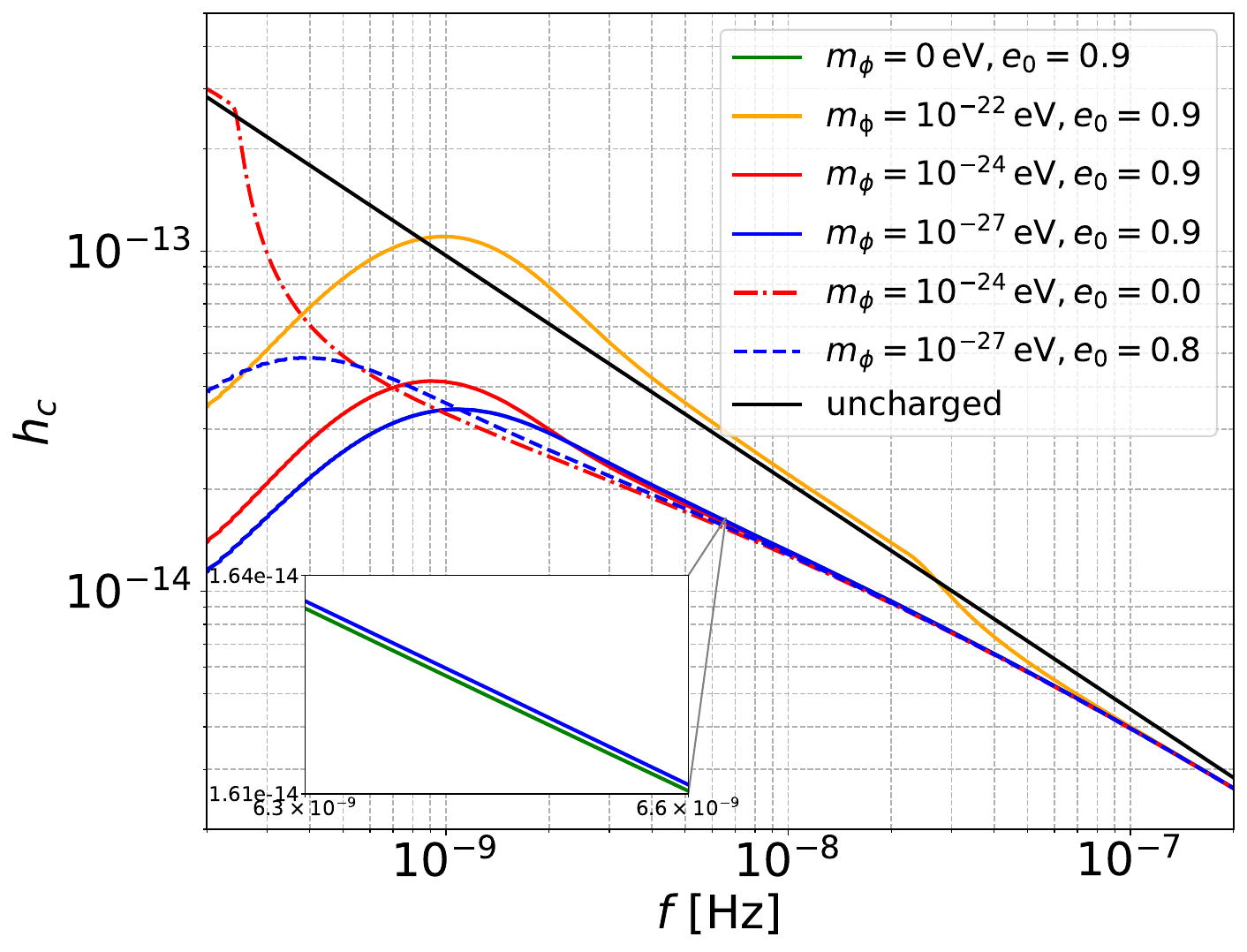}
    \caption{}
    \label{fig:strain_sub2}
  \end{subfigure}

  \vspace{1em}  
  \begin{subfigure}[b]{0.45\textwidth}
    \includegraphics[width=\textwidth]{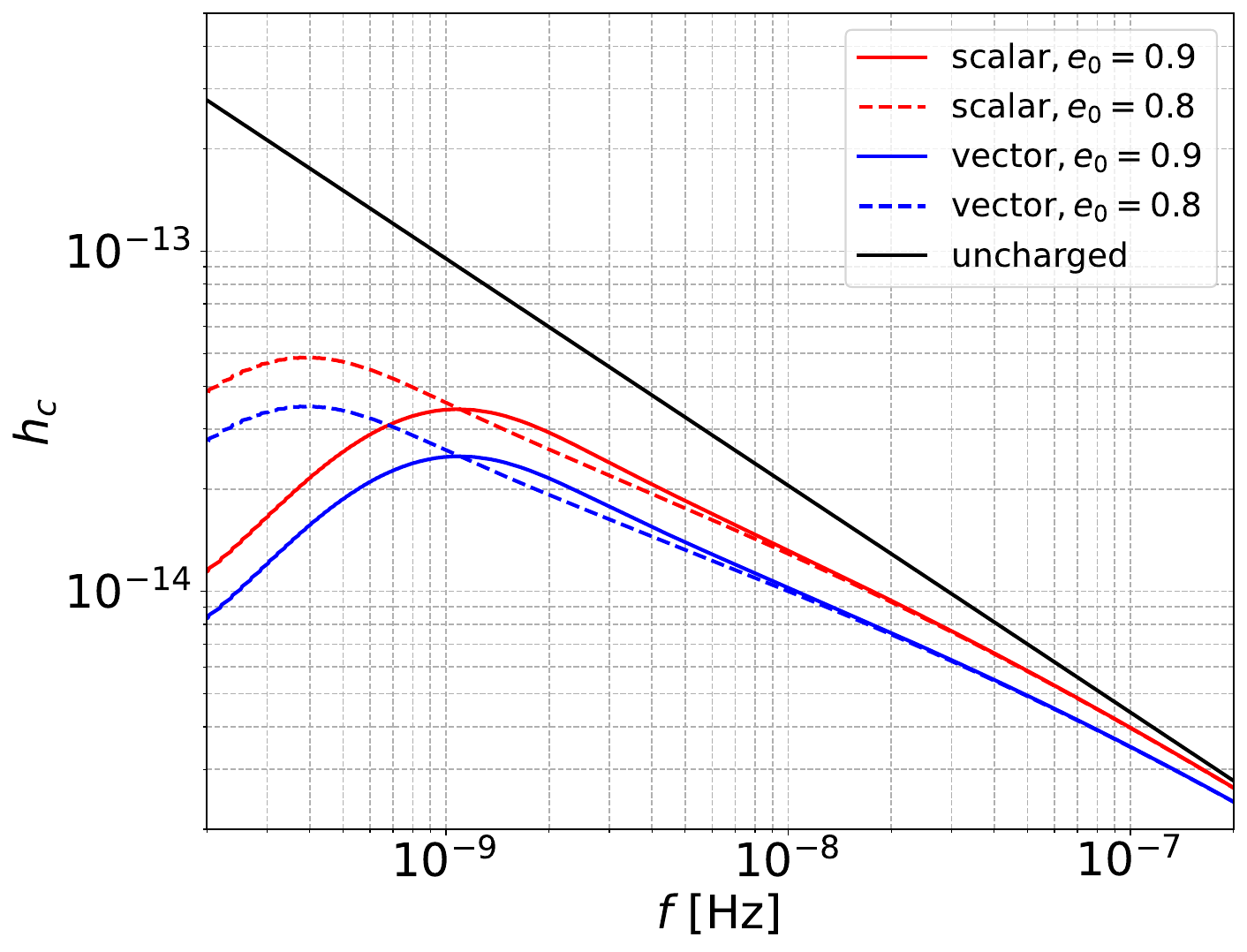}
    \caption{}
    \label{fig:strain_sub3}
  \end{subfigure}%
  \qquad
  \begin{subfigure}[b]{0.45\textwidth}
    \includegraphics[width=\textwidth]{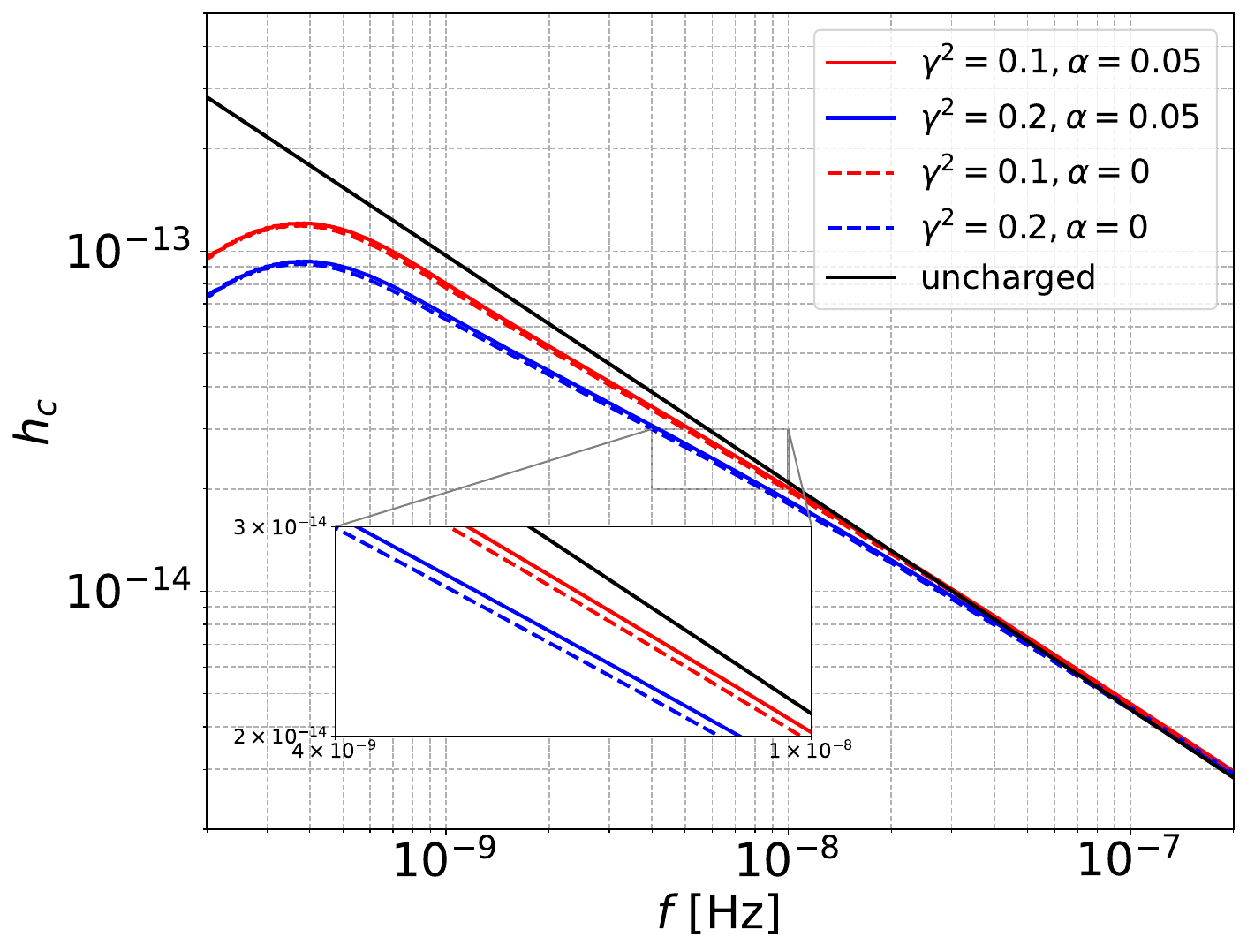}
    \caption{}
    \label{fig:strain_sub4}
  \end{subfigure}

  \jcaption{
  Characteristic strain spectrum for a single source population of SMBHBs with $M = 10^{10}\,M_{\odot}$, $q = 0.1$, and $z = 1$. Panels (a)--(d) illustrate the effects of varying initial eccentricity $e_0$, dipole strength $\gamma$, boson mass $m$, and coupling strength $\alpha$. The total merger rate $N$ is chosen such that $h_c=9.685\times 10^{-15} (f/\mathrm{yr}^{-1})^{-2/3}$ for $e_0=0$ in the uncharged case (black solid line). In panels (a)--(c), we choose $\alpha=0.001$.
  }
  \label{fig:strain}
\end{figure}

Figure~\ref{fig:strain} presents the characteristic strain spectrum for a single source population, corresponding to $\mathrm{d}^3n/(\mathrm{d}z\,\mathrm{d}M\,\mathrm{d}q)=N\,\delta(z-1)\,\delta(M-10^{10}M_\odot)\,\delta(q-0.1)$, with $N$ being the total merger rate. For $e_0>0$, the spectrum develops a peak. As shown in Fig.~\ref{fig:strain_sub1}, for $m_\phi=10^{-25}\,\text{eV}$, increasing $\gamma^2$ reduces the relative contribution of gravitational radiation to the total energy loss, lowering the spectrum while leaving the peak position unchanged. A higher initial eccentricity shifts this peak toward higher frequencies. Since the dipole strength $\gamma$ under consideration is relatively large, the coupling strength $\alpha$ has only a weak influence, as illustrated in Fig.~\ref{fig:strain_sub4}, for $m_\phi=10^{-25}\,\text{eV}$. In the scalar case, since $|\mathrm{d}e/\mathrm{d}\Omega|$ for given $\{\Omega,e\}$, and $\Omega(e=e_0)$ are both increased by $\alpha$, the eccentricity at a given orbital frequency decreases as $\alpha$ increases. Combined with the $\tilde M$-dependence of Eq.~\eqref{dE/df_formula}, increasing $\alpha$ still enhances the spectrum.

The radiation of a lighter field turns on earlier during the inspiral and thus suppresses $h_c(f)$ more strongly, as shown in Fig.~\ref{fig:strain_sub2} for $\gamma^2=0.8$. As $m$ decreases, the peak shifts to higher frequencies (partially mimicking the effect of a larger initial eccentricity), smoothly approaching the massless limit. For the systems considered, the spectrum is no longer sensitive to the boson mass for $m \lesssim 10^{-27}\,\mathrm{eV}$. For circular binaries, the spectrum follows a pure power-law form at $(1+z)f < m/2\pi$, where the dipole radiation is absent. At frequencies above this threshold, $P_{\rm gw}$ gradually dominates over $P_{\rm dip}$, and the spectrum once again approaches the curve in the uncharged case. Figure~\ref{fig:strain_sub3} compares scalar and vector cases for $\gamma^2 = 0.8$ and $m=10^{-25}\,\text{eV}$. Since $P_{\rm dip}$ is stronger for the vector field, it leads to a correspondingly greater suppression of the spectrum.

\begin{figure}[t]
  \centering
  \begin{subfigure}[b]{0.45\textwidth}
    \centering
    \includegraphics[width=\textwidth]{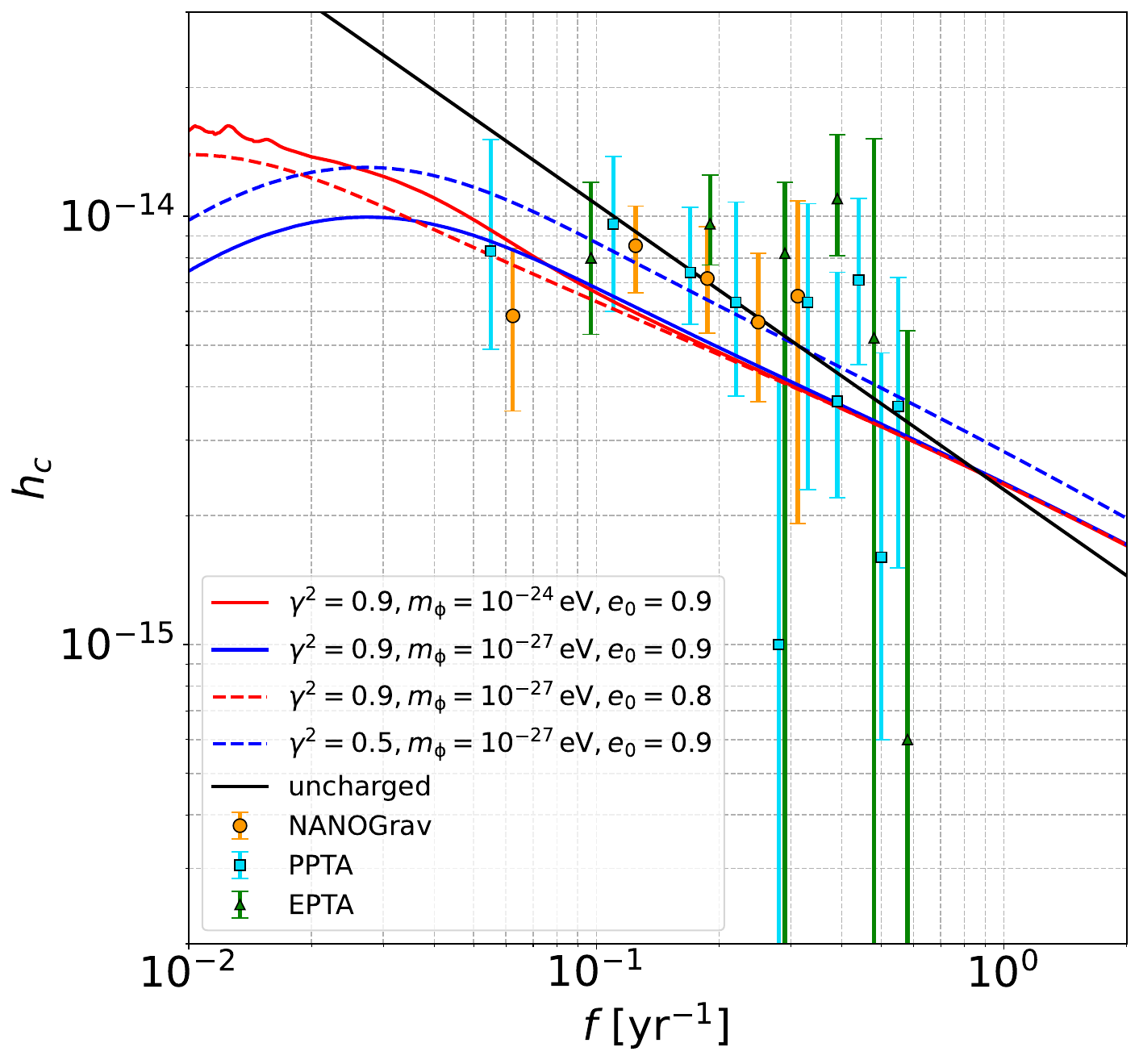}
    \caption{}
    \label{fig:strain_integrate}
  \end{subfigure}
  \qquad
  \begin{subfigure}[b]{0.45\textwidth}
    \centering
    \includegraphics[width=\textwidth]{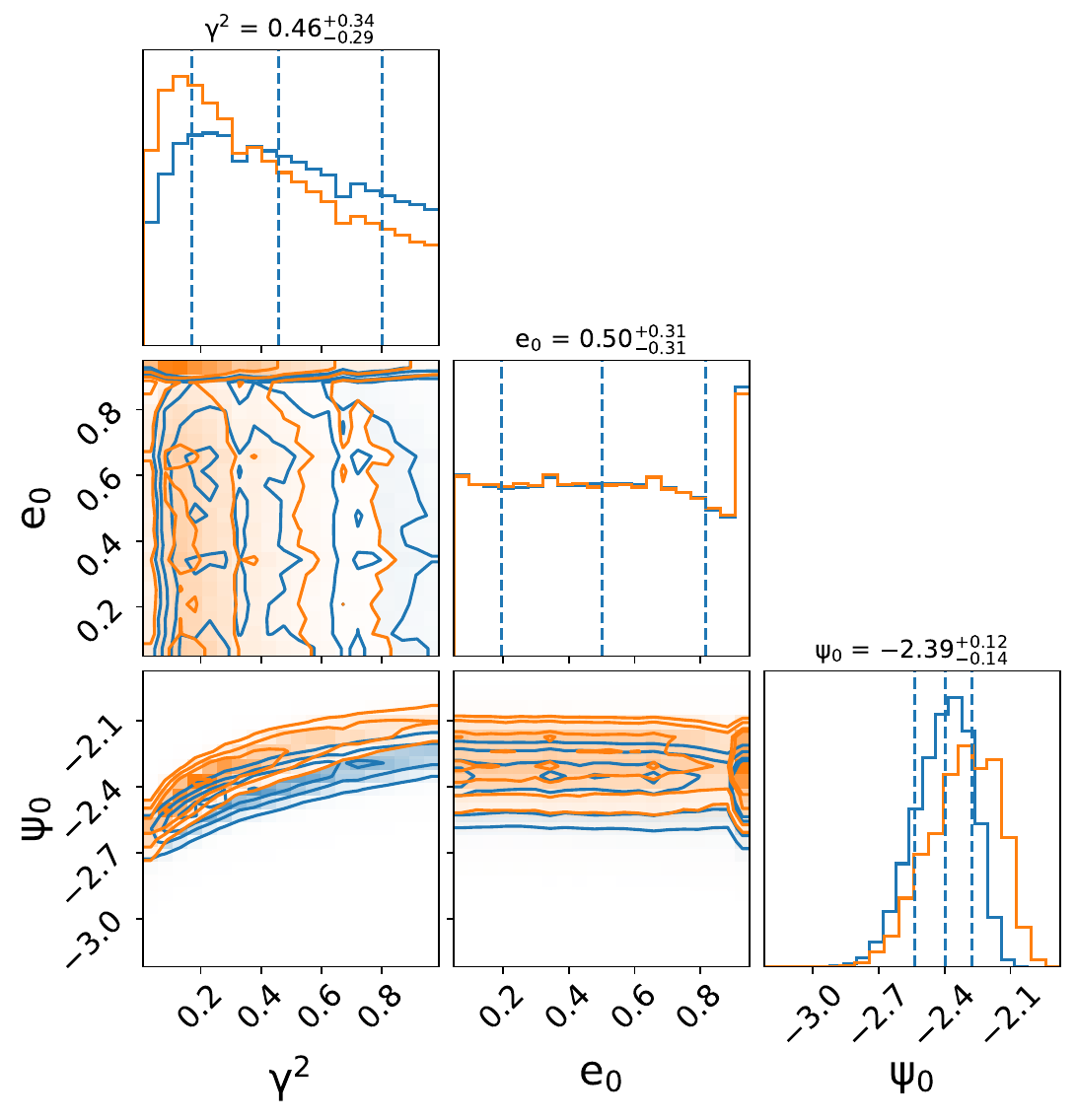}
    \caption{}
    \label{fig:post_distri}
  \end{subfigure}
  \jcaption{
  \textbf{Left:} Characteristic strain spectrum for a representative population of SMBHBs using posterior-median parameters, overlaid with PTA data. The black line corresponds to the uncharged case with $e_0=0$ and $h_c(f) \approx 1.793\times10^{-15}\,\bigl(f/\mathrm{yr}^{-1}\bigr)^{-2/3}$. \textbf{Right:} Posterior distributions of the model parameters for $m=10^{-27}\,\mathrm{eV}$. Results for the scalar field are shown in blue, with posterior medians and $1\sigma$ intervals: $\gamma^2 = 0.46^{+0.34}_{-0.29}$, $e_0 = 0.50^{+0.31}_{-0.31}$, $\psi_0 = -2.39^{+0.12}_{-0.14}$. Results for the vector field are shown in orange, with posterior median and $1\sigma$ intervals: $\gamma^2 = 0.37^{+0.37}_{-0.25}$, $e_0 = 0.52^{+0.30}_{-0.32}$, $\psi_0 = -2.29^{+0.15}_{-0.18}$. In both panels, we choose $\alpha=0.001$.
  }
  \label{fig:result}
\end{figure}

Using the merger-rate model described in Appendix~\ref{app:B}, we compute the characteristic strain spectrum for a representative population of SMBHBs, assuming for simplicity that each binary has the same eccentricity $e_0$ at $a=1\,\text{pc}$~\cite{chen2025galaxytomographygravitationalwave}. The results are shown in Fig.~\ref{fig:strain_integrate}. Notably, the spectra for $m_\phi=10^{-24}\,\text{eV}$ and $m_\phi=10^{-27}\,\text{eV}$ are clearly distinguishable at low frequencies, which could be accessible to future observations. Using the lowest five frequency bins of the NANOGrav 15-year data, we find a minimum chi-squared value of $\chi^2=1.189$ for our model incorporating dipole radiation, which is much smaller than the corresponding value of $\chi^2=6.445$ for the uncharged model with $e_0=0$.

To assess the consistency of this model with PTA datasets, we perform a Bayesian analysis of the parameters $\boldsymbol{\Theta}=\{\gamma^2, e_0, \psi_0\}$, where $\psi_0$ is the normalization parameter for the merger rate. We assign a normal prior $P(\psi_0)=\mathcal{N}[-2.56,0.4]$ to $\psi_0$ and uniform priors over $[0, 1]$ to both $\gamma^2$ and $e_0$. The likelihood function is taken to be
\begin{equation}\label{eq:likelyhood}
\ln\mathcal{L}(\boldsymbol{\Theta}) = \sum_i -\frac{1}{2}\left[ \frac{h_{c,i}-h_c(f_i;\boldsymbol{\Theta})}{\sigma_i} \right]^2,
\end{equation}
where $h_c(f_i;\boldsymbol{\Theta})$ is the model prediction, and $h_{c,i}$ and $\sigma_i$ denote the central value and the mean of the upper and lower uncertainties of the free-spectrum data at the frequency bin centered at $f_i$, as presented in Table~\ref{Tab:pta_hc_values}. The posterior distribution is given by
\begin{equation}
P\left(\boldsymbol{\Theta}|h_{c,i}\right)=\frac{P\left(h_{c,i}|\boldsymbol{\Theta}\right)P\left(\boldsymbol{\Theta}\right)}{P\left(h_{c,i}\right)}\propto\mathcal{L}(\boldsymbol{\Theta})\,P(\boldsymbol{\Theta}).
\end{equation}

\begin{figure}[hbt!]
  \centering
  \begin{subfigure}[b]{0.45\textwidth}
    \centering
    \includegraphics[width=\textwidth]{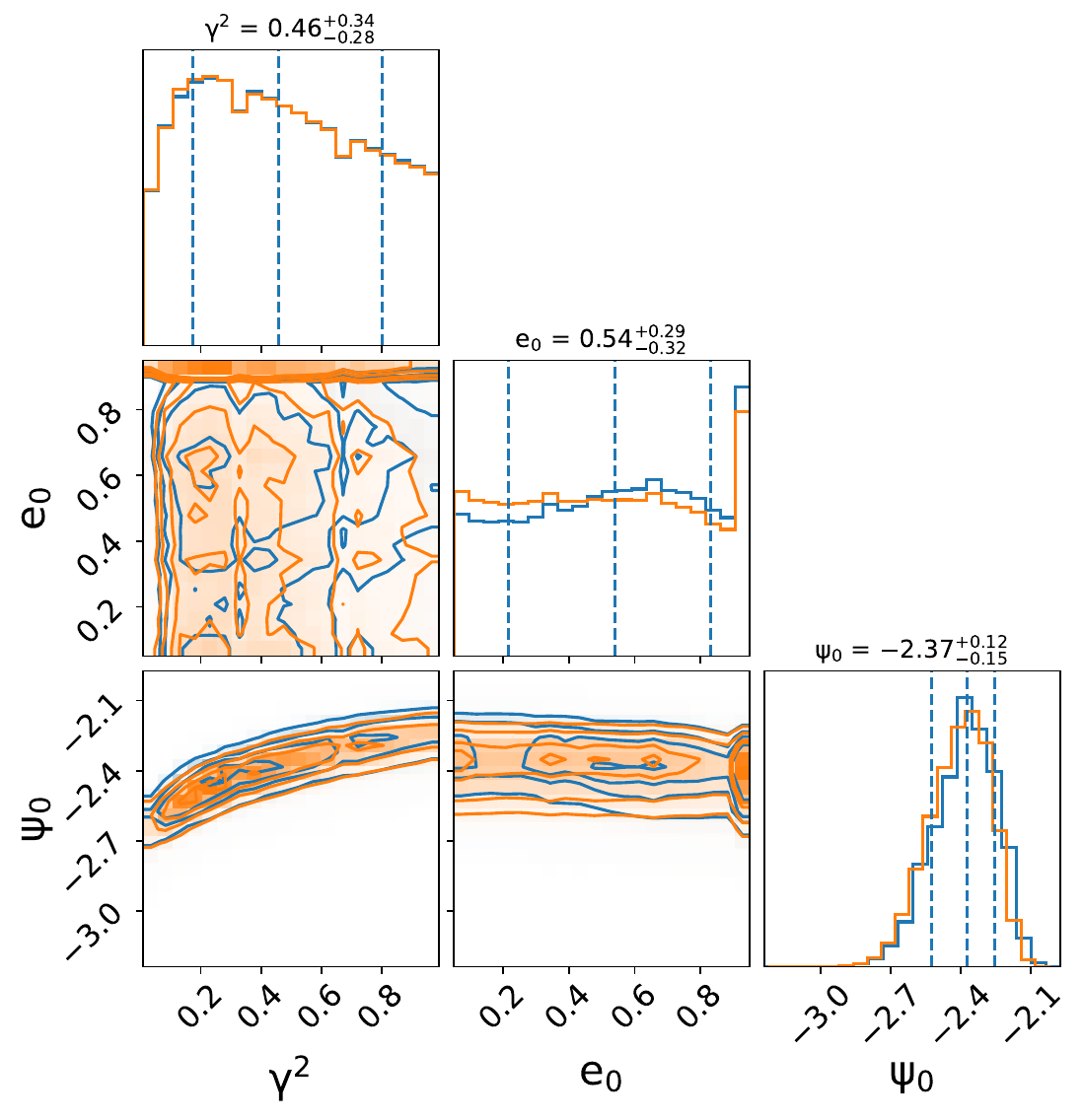}
    \caption{}
    \label{fig:post_distri_scalar}
  \end{subfigure}
  \qquad
  \begin{subfigure}[b]{0.45\textwidth}
    \centering
    \includegraphics[width=\textwidth]{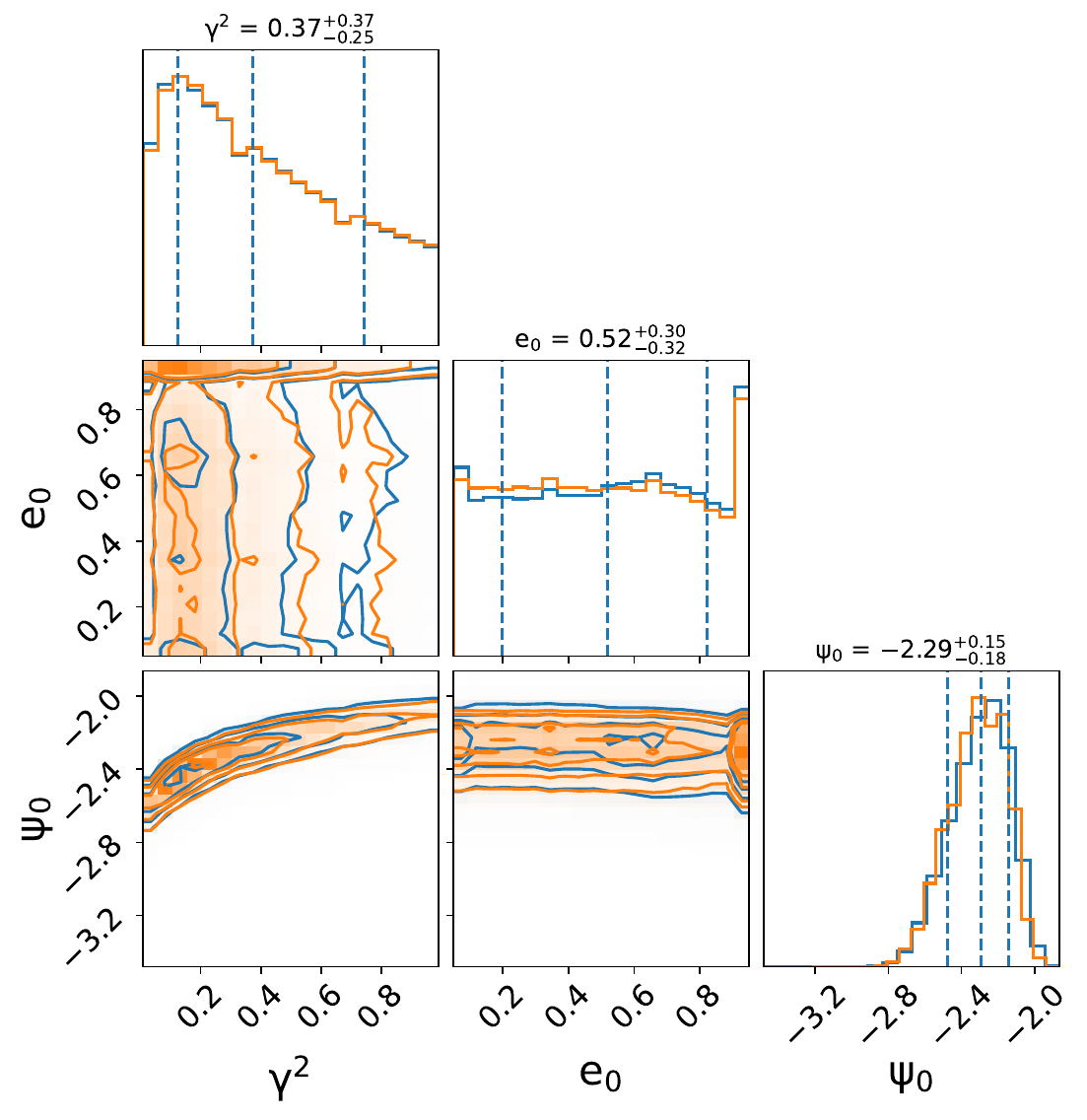}
    \caption{}
    \label{fig:post_distri_vector}
  \end{subfigure}
  \jcaption{
  Posterior distributions in the scalar case (left) and the vector case (right). Results for $m=10^{-27}\,\text{eV}$ ($10^{-25}\,\text{eV}$) are shown in blue (orange). In both panels, we choose $\alpha=0.001$.
  }
  \label{fig:post_distri_all}
\end{figure}

Figures~\ref{fig:post_distri} and \ref{fig:post_distri_all} display the posterior distributions in the scalar and vector cases with $m\in\{10^{-27},10^{-25}\}\,\text{eV}$. Since the boson mass has a rather limited influence on the spectrum at $f>1\,\text{nHz}$ ($\approx 0.03\,\text{yr}^{-1}\approx 6.6\times 10^{-25}\,\text{eV}$), the posteriors for $m=10^{-27}\,\text{eV}$ and $m=10^{-25}\,\text{eV}$ exhibit only minor differences, primarily reflected in shifts of the median of the normalization parameter $\psi_0$. For the same boson mass, the posteriors in the scalar and vector cases show more pronounced distinctions. Compared with the scalar case, the posterior in the vector case favors smaller values of $\gamma^2$ and is more narrowly concentrated; it also favors a larger normalization parameter $\psi_0$, indicating a preference for higher galaxy merger rates. The posterior of the initial eccentricity remains close to its uniform prior, with only a slight increase in probability density at very high eccentricities. The posterior distributions of $\gamma^2$ and $\psi_0$ exhibit well-defined peaks, whereas no clear preference is observed for $e_0$. The peak of $\psi_0$ increases as the boson mass decreases. This behavior can be understood from the $\psi_0$-dependence of the binary merger rate $\mathrm{d}^3n/(\mathrm{d}z\,\mathrm{d}M\,\mathrm{d}q) \propto 10^{\psi_0}$~\cite{2023ApJ...952L..37A}. A lighter boson mass leads to stronger suppression of the spectrum, which in turn results in an increase of $\psi_0$ to compensate, as reflected in its posterior distribution.

\section{Conclusion}
\label{sec:conclusion}

We have investigated the effect of dark dipole radiation on the orbital evolution and the stochastic gravitational wave background of supermassive black hole binaries carrying scalar or vector charges within the Newtonian-order Keplerian approximation. Dipole radiation provides an eccentricity-enhanced channel that accelerates the inspirals of SMBHBs, although it alone does not offer a universal solution to the final-parsec problem. Nonetheless, the SGWB from SMBHB inspirals can be influenced by the presence of dipole radiation. Bayesian fits of our simplified SGWB model to the PTA data appear to favor a nonzero dipole strength, while the posterior of the initial eccentricity remains broadly consistent with a uniform prior and shows mild support for very high eccentricities. The current PTA data are also insensitive to the boson mass $m\lesssim 10^{-25}\,\text{eV}$, given its relatively small influence on the spectrum within the observational frequency band. Future measurements extending to $f \lesssim 1\,\mathrm{nHz}$ have the potential to probe the parameter space more thoroughly.

\begin{acknowledgments}
We would like to thank Jun Zhang, Yifan Chen, and Yong Tang for useful discussions and valuable input. Y. C. is grateful to Ya-Ze Cheng and Wen-Hao Wu for earlier discussions on this topic. We also thank the anonymous referee for valuable comments and suggestions that helped improve the quality of the manuscript.
\end{acknowledgments}

\appendix
\section{Conserved charge densities of scalar and vector fields}
\label{App1}

\subsection{Scalar field}
The Lagrangian of a free real scalar field in flat spacetime is
\begin{equation}
\begin{aligned}
\mathcal{L}_\phi &= \frac{1}{2} \partial_a \phi\,\partial^a\phi-\frac{1}{2}m_\phi^2 \phi^2
=
\frac{1}{2} \left[\dot\phi^2-(\partial_i \phi)(\partial_i \phi)\right]-\frac{1}{2}m_\phi^2 \phi^2.
\end{aligned}
\end{equation}
The EMT is
\begin{equation}
T_{ab}=\partial_a\phi\,\partial_b\phi -\eta_{ab}\mathcal{L}_\phi=\partial_a\phi\,\partial_b\phi- \eta_{ab}\left(\frac{1}{2}\partial_c\phi\,\partial^c\phi-\frac{1}{2}m_\phi^2\phi^2\right),
\end{equation}
with components
\begin{align}
T_{00} =\frac{1}{2}\left[\dot\phi^2+(\partial_i\phi)(\partial_i\phi)\right]+\frac{1}{2}m_\phi^2\phi^2,
\quad
T_{0i} =\dot\phi\,\partial_i\phi.
\end{align}

The coordinate transformation corresponding to an infinitesimal spatial rotation about $\mathbf{x=0}$ is $x^i\to  x^i+\omega_{ij}x^j=x^i+\sum_{k<l}\omega_{kl}A_{kl}^i$, with $\omega_{ij}=-\omega_{ji}$ and $A^i_{kl}\equiv \delta_{ik}x^l-\delta_{il}x^k$. By Noether's first theorem (see for example \cite{GW_volume_1}), the invariance of the action under global spatial rotations is associated with the conserved charge $J_{kl}=\int d^3x\,j_{kl}$, with the local charge density
\begin{equation}
\begin{aligned}
j_{kl} =\frac{\partial \mathcal{L}_\phi}{\partial \dot \phi}A^j _{kl}\partial_j \phi
=\dot\phi \left(x^l\partial_k\phi-x^k\partial_l\phi\right)
.
\end{aligned}
\end{equation}
This gives the angular momentum density
\begin{equation}
\begin{aligned}
j_s =\frac{1}{2}\epsilon_{skl}\,j_{kl}
=\epsilon_{skl}x^k\left(-\dot\phi \partial_l\phi\right).
\label{j_scalar}
\end{aligned}
\end{equation}

\newpage
\subsection{Vector field}
The Lagrangian of a free real vector field in flat spacetime is
\begin{equation}
\begin{aligned}
\mathcal{L}_A &= -\frac{1}{4} F_{ab}F^{ab} +\frac{1}{2}m_A^2 A_a A^a
\\
&=
\frac{1}{2} \left[\dot A_i \dot A_i+(\partial_i A_0)(\partial_i A_0)-(\partial_i A_j)(\partial_i A_j)-2\dot A_i(\partial_i A_0)+(\partial_i A_j)(\partial_j A_i)\right] +\frac{1}{2}m_A^2 (A_0^2-A_iA_i)
.
\end{aligned}
\end{equation}
Note that $A_0=A^0$ and $A_i=-A^i$. The EMT is
\begin{equation}
T_{ab}=-{F_a}^dF_{bd}+m_A^2A_aA_b-\eta_{ab}\mathcal{L}_A=\frac{1}{4}\eta_{ab}F_{cd}F^{cd}-\eta^{cd}F_{ac}F_{bd}+m^2_A\left(A_aA_b-\frac{1}{2}\eta_{ab}A_c A^c\right),
\end{equation}
with components
\begin{align}
T_{00} 
&=\frac{1}{2} \left[\dot A_i \dot A_i+(\partial_i A_0)(\partial_i A_0)+(\partial_i A_j)(\partial_i A_j)-2\dot A_i(\partial_i A_0)-(\partial_i A_j)(\partial_j A_i)\right] +\frac{1}{2}m_A^2 (A_0^2+A_iA_i)
,
\\
T_{0i} &=(\dot A_j-\partial_j A_0)(\partial_i A_j-\partial_j A_i)+m^2_A A_0A_i.
\end{align}

Under an infinitesimal spatial rotation $x^i \to x^i+\omega_{ij}x^j$, the spatial components of the vector field transform as $A_{i}\to (\delta_{ij}+\omega_{ij})A_j=A_i+\sum_{k<l}\omega_{kl}F_{i,kl}$, where $F_{i,kl}\equiv\delta_{ik}A_l-\delta_{il}A_k$. Consequently, the angular momentum density is given by\footnote{Note that Eqs.~\eqref{j_scalar} and \eqref{j_vector} also hold for interacting fields, provided that the interaction Lagrangian preserves rotational symmetry and does not depend explicitly on $\dot\phi$ or $\dot A_a$.}
\begin{equation}
\begin{aligned}
j_s &=\frac{1}{2}\epsilon_{skl}\left[\frac{\partial \mathcal{L}_A}{\partial \dot A_i}(A^j _{kl}\partial_j A_i-F_{i,kl})
\right]
=\epsilon_{skl}\left(\dot A_i-\partial_i A_0\right)\left(-x^k\partial_l A_i-\delta_{ik}A_l\right)
,\label{j_vector}
\end{aligned}
\end{equation}
which satisfies $\int d^3x\,j_i=\int d^3x\, \epsilon_{ijk}x^jT^{0k}$, up to a boundary term.

For $m_A=0$, choosing $A_0=0$ gives the result for a massless Abelian vector field in the temporal gauge. We can also check the nonrelativistic (NR) limit of this result for $m_A>0$. The real massive vector field can be represented by $A_i=\frac{1}{\sqrt{2m_A}}(\Psi_i\,e^{-im_A t} + \text{c.c.} )$. In the NR limit ($|\partial_iA_a|\ll m_A|A_a|\sim |\partial_t A_a|$), the slow mode is described by the wavefunction $\Psi_i$, with $|A_0|\ll |\mathbf{A}|$. The EMT is dominated by $T_{00}$ and $T_{ij}$, while
\begin{equation}
\begin{aligned}
j_s \approx -\epsilon_{skl} (x^k \dot A_i \partial_l A_i+\dot A_k A_l)
\approx -i\epsilon_{skl} (x^k \Psi_i^* \partial_l \Psi_i+ \Psi_k^* \Psi_l)
,
\end{aligned}
\end{equation}
The first and second terms correspond, respectively, to the orbital and spin\footnote{In the spherical basis, $\boldsymbol{\Psi}=\sum_{i=x,y,z}\Psi_i \mathbf{e}_i =\sum_{m=1,0,-1}\psi_m \boldsymbol{\xi}^m$ with $\boldsymbol{\xi}^0=\mathbf{e}_z$ and $\boldsymbol{\xi}^{\pm 1}=\mp (\mathbf{e}_x\pm i \mathbf{e}_y)/\sqrt{2}$, the spin angular momentum density reads $\mathbf{j}_\text{spin}=i\bold{\Psi}\times \bold{\Psi}^*=\psi_m^* \hat{\mathbf{S}}_{mn} \psi_n$, where $\hat{\mathbf{S}}$ is the spin operator with $\hat S_z=\text{diag}(1,0,-1)$.} contributions, with $-i\epsilon_{skl}=(\hat S_s)_{kl}$ being the spin operator in the adjoint representation of SU(2); this, indeed, coincides with the angular momentum density in the NR theory~\cite{Jain_2022}. The energy flux density of the vector field in the NR limit is
\begin{equation}
p_i=-T_{0i}\approx -\dot A_j(\partial_i A_j-\partial_j A_i)-m_A^2A_0A_i\approx \text{Im}[\Psi_j^*\partial_i \Psi_j-\partial_j(\Psi_j^*\Psi_i)],
\end{equation}
(note that here $A_0$ cannot be neglected) which is also the conserved current associated with the global U(1) symmetry of the NR theory, with the conserved charge density given by $m_A\Psi^*_i\Psi_i\approx T_{00}$. The second part $p_i^\text{spin}=-\text{Im}[\partial_j(\Psi_j^*\Psi_i)]$ is automatically divergence-free and is related to the spin angular momentum since (up to a boundary term)
\begin{equation}
\int d^3x\,\mathbf{x}\times \mathbf{p}^\text{spin}=
-\mathbf{e}_k\,\epsilon_{kji}\int d^3x\,x^j\,\text{Im}[\partial_l(\Psi_l^*\Psi_i)]
=
\mathbf{e}_k\int d^3x\,\left(i\,\epsilon_{kil}\,\Psi_i\Psi_l^*\right)
=
\int d^3x\,\mathbf{j}_\text{spin}
.
\end{equation}
$p_i^\text{spin}$ is typically neglected in NR theory, but it contributes to the generation of gravitomagnetic fields~\cite{Cao:2024wby}. In the NR limit, free on-shell bosonic fields behave as homogeneous, pressureless dust under time averaging~\cite{Yu:2024enm}. However, if the fields form a gravitationally bound structure, they can no longer be treated as free, radiative fields.

\section{Merger rate of SMBHBs}
\label{app:B}
The merger rate of SMBHBs is a function of a redshift, mass ratio, and total mass, which can be related to the galaxy merger rate via~\cite{2023ApJ...952L..37A,alonsoálvarez2024selfinteractingdarkmattersolves},
\begin{equation}\label{eq:event_density}
    \frac{\mathrm{d}^3n}{\mathrm{d}z\,\mathrm{d}M\,\mathrm{d}q}=\frac{\mathrm{d}^3n_g}{\mathrm{d}z\,\mathrm{d}M_{\star}\,\mathrm{d}q_{\star}}\,\frac{\mathrm{d}M_{\star}}{\mathrm{d}M}\,\frac{\mathrm{d}q_{\star}}{\mathrm{d}q},
\end{equation}
where $M_{\star}$ denotes the stellar mass of the more massive galaxy in the initial merger, and $q_{\star}$ denotes the mass ratio of the two galaxies. Here, we neglect the effects of dark charges on the merger rate distribution. We use the following relations between the SMBHB total mass $M$, the bulge mass $M_{\mathrm{bulge}}$ and stellar mass $M_{\star}$ of the host galaxy~\cite{Kormendy_2013,Chen_2019}:
\begin{align}
	\mathrm{log}_{10}\left(\frac{M}{\,M_{\odot}}\right) &=8.7+1.1\,\mathrm{log}_{10}\left(\frac{M_{\rm bulge}}{10^{11}\,M_{\odot}}\right), \label{eq:Mtot_Mbulge}
	\\
	M_{\rm bulge}&=(0.615 + df_{\star})M_{\star}, \label{eq:Mbulge_Mstar}
\end{align}
with
\begin{equation}
	df_{\star} = 
	\begin{cases}
		0, & M_{\star} \leq 10^{10}\,M_{\odot} \\
		\frac{\sqrt{6.9} \exp \left( \frac{-3.45}{\log_{10} M_{\star} - 10} \right)}{(\log_{10} M_{\star} - 10)^{1.5}}, & M_{\star} > 10^{10}\,M_{\odot}
	\end{cases}
\end{equation}
The galaxy merger rate can be expressed as~\cite{2023ApJ...952L..37A,Chen_2019}
\begin{equation}\label{eq:event_density_of_galaxy}
    \frac{\mathrm{d}^3n_g}{\mathrm{d}z\,\mathrm{d}M_\star\,\mathrm{d}q_{\star}}=\frac{\Psi(M_{\star},z')}{M_{\star}}\,\frac{P(M_{\star},q_{\star},z')}{T_\text{g-g}(M_{\star},q_{\star},z')}\,\frac{\mathrm{d}t}{\mathrm{d}z'};
\end{equation}
where $P$ is the galaxy pair function, and $T_\text{g-g}$ is the galaxy merger time (measured in the cosmic time $t$),
\begin{equation}
    \Psi(M_{\star},z)=\Psi_0\left(\frac{M_{\star}}{M_{\Psi}}\right)^{\alpha_{\Psi}}\mathrm{exp}\left(-\frac{M_{\star}}{M_{\Psi}}\right);
\end{equation}
with
\begin{align}
    \log_{10} \left( {\Psi_0}/{\mathrm{Mpc}^{-3}} \right) &= \psi_0 + \psi_z\, z;  \\ \label{eq:psi_constant_def}
    \log_{10} \left( {M_\Psi}/{M_\odot} \right) &= m_{\psi 0} + m_{\psi z}\, z;
    \\
    \alpha_\Psi &= 1 + \alpha_{\psi 0} + \alpha_{\psi z}\, z.
\end{align}
The galaxy pair function and merger time can be modeled by
\begin{align}
    P\left(M_{\star},q_{\star},z\right) =P_0(1+z)^{\beta_{p0}};
\quad
    T_\text{g-g}(M_{\star},q_{\star},z) =T_0(1+z)^{\beta_{t0}}q^{\gamma_{t0}}_{\star}.
\end{align}
The values of parameters in these polynomials are listed in Table~\ref{Tab:density_parameters}. The redshift $z'$ in Eq.~\eqref{eq:event_density_of_galaxy} corresponds to the onset of a galaxy merger, thus
\begin{align}
    t(z)-t(z')=T_\text{g-g}(z'),\quad
    t(z=0)=13.79~\mathrm{Gyr};
\end{align}
where $t(z)$ is given by the $\Lambda$CDM model
\begin{align}
    \frac{\mathrm{d}t}{\mathrm{d}z}=-\frac{1}{(1+z)\,H(z)};
    \quad
    H(z)=H_0\sqrt{\Omega_{\Lambda}+(1+z)^3\,\Omega_m};
\end{align}
and we use $H_0=67.4\,\mathrm{km}\,\mathrm{s}^{-1}\mathrm{Mpc}^{-1}$, $\Omega_m=0.315$, and $\Omega_{\Lambda}=0.685$~\cite{2020}.

\begin{table}[htb]
    \caption{Values of parameters used in the merger-rate model~\cite{alonsoálvarez2024selfinteractingdarkmattersolves}.}
    \label{Tab:density_parameters}
        \begin{tabular}{cc|cc}
         \hline
         \hline
            \textbf{ Parameter } & \textbf{ Value } & \textbf{ Parameter } & \textbf{ Value } \\
            \hline
            $\psi_0$ & $\mathrm{Free}$ & $P_0$ & $0.033$ \\
            $\psi_z$ & $-0.6$ & $\beta_{p0}$ & $1$ \\
            $m_{\psi 0}$ & $11.5$ & $T_0$ & $0.5 \, \mathrm{Gyr}$ \\
            $m_{\psi z}$ & $0.11$ & $\beta_{t0}$ & $-0.5$ \\
            $\alpha_{\psi 0}$ & $-1.21$ & $\gamma_{t0}$ & $-1$ \\
            $\alpha_{\psi z}$ & $-0.03$ & & \\
        \hline
        \hline
        \end{tabular}
\end{table}

\begin{figure}[hbt!]
  \centering
  \begin{subfigure}[b]{0.5\textwidth}
    \centering
    \includegraphics[width=\textwidth]{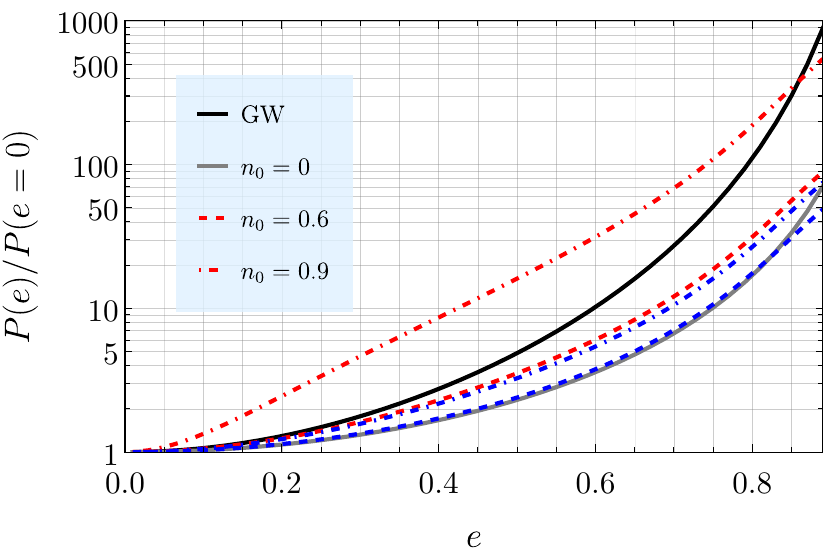}
    \caption{}
    \label{fig:P}
  \end{subfigure}
  \qquad
  \begin{subfigure}[b]{0.5\textwidth}
    \centering
    \includegraphics[width=\textwidth]{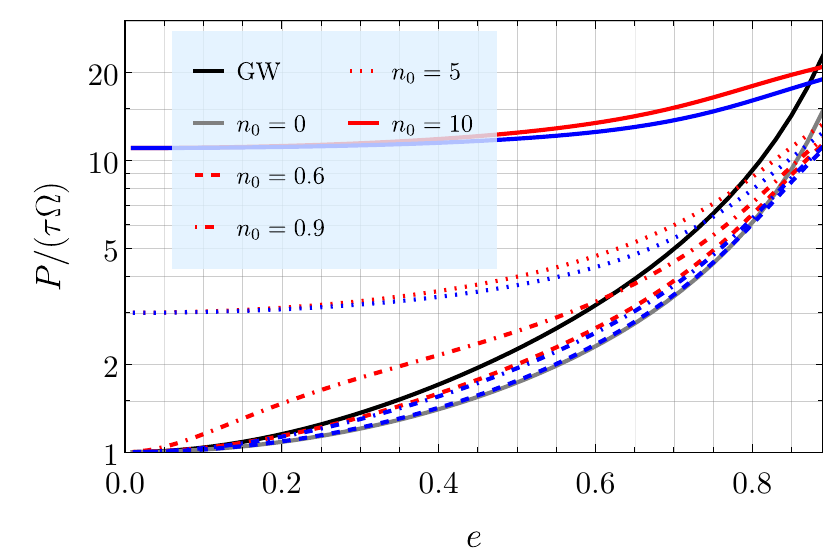}
    \caption{}
    \label{fig:P/tau}
  \end{subfigure}
  \jcaption{
  Enhancement of the energy flux with orbital eccentricity shown as the ratio $P(e)/P(e=0)$ (top panel) and the ratio $P/(\tau\Omega)$ (bottom panel) for the gravitational quadrupole radiation (black), scalar (red), and vector (blue) dipole radiation. The massless case corresponds to $n_0=m/\Omega=0$.
  }
\end{figure}

\setlength{\tabcolsep}{8pt}
\begin{table}[htb]
\centering
\caption{Central values and uncertainties of the PTA free-spectrum data.}
\label{Tab:pta_hc_values}
{\renewcommand{\arraystretch}{1.35}
\begin{tabular}{cc|cc}
\hline\hline
$f\,[\mathrm{yr}^{-1}]$ & $h_c/10^{-15}$ & $f\,[\mathrm{yr}^{-1}]$ & $h_c/10^{-15}$ \\
\hline
\multicolumn{2}{c}{\textbf{NANOGrav}~\cite{2023ApJ...952L..37A}} & \multicolumn{2}{c}{\textbf{PPTA}~\cite{antoniadis2024seconddatareleaseeuropean}} \\
0.062 & $6.5^{+4.5}_{-2.4}$ & 0.055 & $8.3^{+6.8}_{-3.4}$ \\
0.12  & $7.9^{+3.0}_{-1.8}$ & 0.11  & $9.6^{+4.2}_{-3.6}$ \\
0.19  & $7.4^{+3.1}_{-2.0}$ & 0.17  & $7.4^{+3.1}_{-1.8}$ \\
0.25  & $6.4^{+3.3}_{-1.9}$ & 0.22  & $6.3^{+4.5}_{-2.5}$ \\
0.31  & $9.3^{+4.8}_{-4.2}$ & 0.28  & $1.0^{+3.2}_{-0.8}$ \\
\hline
\multicolumn{4}{c}{\textbf{EPTA}~\cite{Reardon_2023}} \\
0.097 & $8.0^{+4.0}_{-2.7}$ & 0.33 & $6.3^{+4.4}_{-4.0}$ \\
0.19  & $9.6^{+2.9}_{-1.9}$ & 0.39 & $3.7^{+3.7}_{-1.5}$ \\
0.29  & $8.2^{+3.8}_{-2.8}$ & 0.44 & $7.1^{+3.9}_{-2.6}$ \\
0.39  & $6.1^{+4.5}_{-2.9}$ & 0.50 & $1.6^{+3.2}_{-1.0}$ \\
0.48  & $10.1^{+10.0}_{-5.5}$ & 0.55 & $3.6^{+3.6}_{-2.09}$ \\
0.58  & $0.6^{+4.8}_{-0.4}$  &      &                         \\
\hline\hline
\end{tabular}
}
\end{table}

\clearpage
\bibliographystyle{apsrev4-2}
\bibliography{paper}

\clearpage
\begingroup
\renewcommand{\theequation}{\arabic{equation}}
\renewcommand{\theHequation}{erratum.\arabic{equation}}
\setcounter{equation}{0}
\setcounter{footnote}{0}
\hypersetup{linkcolor=blue}
\renewcommand{\addcontentsline}[3]{}

\newcommand{\avg}[1]{\left\langle #1\right\rangle}
\newcommand{\intk}{\int\!\frac{d^3k}{(2\pi)^3}}
\newcommand{\ii}{\mathrm{i}}
\newcommand{\eps}{\epsilon}
\begin{bibunit}[apsrev4-2]

\pdfbookmark[0]{Erratum}{erratum}
\begin{center}
{\Large\bfseries Erratum\par}
\end{center}

Equation~(3) in this paper omitted the factor that arises when the proper-time measure in the covariant scalar worldline coupling of Eq.~(1) is expressed in coordinate time. The correction first enters beyond Newtonian order. For a generic noncircular orbit, it produces a time-dependent scalar monopole at the next order in the source expansion. The term $-2\dot A_i\partial_iA_0$ was mistakenly omitted from the square brackets in Eq.~(22). A sign was also missed in the first term of the second line of Eq.~(25), and another term was incorrectly discarded in passing to the third line. Consequently, the first term in the Proca angular-momentum flux of Eq.~(26) must change sign, and one additional term must be included. We correct these points below. The Newtonian-order dipole fluxes, and hence the calculations and results based on them, are unchanged. The flux formulas in this paper were derived using a particular choice of Noether currents; here we also provide a source-side derivation of the results.

We consider the scalar (Klein-Gordon) and vector (Proca) cases separately, so the subscripts $\{\phi,A\}$ will be omitted in each case when there is no ambiguity. We use $\eta_{ab}=\mathrm{diag}(1,-1,-1,-1)$, $\square=-\partial_t^2+\partial_i\partial_i$, and $\hbar=c=G=1$. $\{n_i,F_i,\tau_i\}$ denote the Cartesian components of $\{\mathbf n,\mathbf F,\boldsymbol\tau\}$, with $|\mathbf n|=1$; $\mathbf F$ and $\boldsymbol\tau$ are time-averaged radiation fluxes. All other conventions remain unchanged.

\section*{I. Correction to the scalar source}

The scalar part of the worldline action in Eq.~(1) is $ S_{\mathrm{int},\phi} =\sum_I q_{I\phi}\int ds_I\,\phi(X_I)$, where $ds_I=\sqrt{g_{ab}\,dX_I^a dX_I^b}$. In flat spacetime, with $X_I^a=(t,\mathbf{X}_I(t))$, $ds_I=\frac{dt}{\gamma_I}$, where $\gamma_I=\frac{1}{\sqrt{1-\mathbf v_I^2}}$. Consequently, Eq.~(3) should read $(\square-m^2)\phi=-n$, with
\begin{align}
n(t,\mathbf{x})
 =\sum_I\frac{q_{I\phi}}{\gamma_I(t)}
 \delta^3\!\left(\mathbf{x}-\mathbf{X}_I(t)\right).
 \label{erratum:eq:correct-source}
\end{align}
The factor $1/\gamma_I$ in the scalar source was omitted in Eq.~(3). Its PN expansion begins as
\begin{equation}
 n(t,\mathbf{x})
 =\sum_I q_{I\phi}
 \left(1-\frac{|\mathbf v_I|^2}{2}\right)
 \delta^3\!\left(\mathbf{x}-\mathbf{X}_I(t)\right)
 +\mathcal{O}(v^4).
 \label{erratum:eq:source-flat-pn}
\end{equation}
For constant vector charges, the coupling $q_{IA}\int dX_I^a A_a$ gives the current $J^a(t,\mathbf{x})=\sum_I q_{IA}v_I^a\delta^3\!\left(\mathbf{x}-\mathbf{X}_I\right)$, with $v_I^a=(1,\mathbf v_I)$, $\partial_a J^a=0$, and $(\square -m^2)A^a=J^a$.

The $\mathcal{O}(v^2)$ term in Eq.~\eqref{erratum:eq:source-flat-pn} first changes the scalar dipole amplitude at relative 1PN order. It also contributes to the scalar monopole at $\mathcal{O}(v^2)$. For constant scalar charges, the Newtonian monopole $\sum_I q_{I\phi}$ is time independent. Footnote 3 correctly notes that time-dependent scalar charges can generate a time-dependent Newtonian monopole, but even constant charges generically produce a time-dependent monopole source at $\mathcal{O}(v^2)$ on a noncircular orbit. The Newtonian-order dipole results are unchanged.

\section*{II. Quadrupole-order approximation}
For the multipole expansion, it is convenient to extend the harmonic source in Eq.~(11) off shell and define
\begin{equation}
\begin{aligned}
 n_{(n)}(\mathbf{k})
 &\equiv\frac{1}{T}\int_0^Tdt\,e^{\ii\Omega_n t}
 \sum_I\frac{q_{I\phi}}{\gamma_I(t)}
 e^{-\ii\mathbf{k}\cdot\mathbf{X}_I(t)},
\end{aligned}
 \label{erratum:eq:nn-offshell}
\end{equation}
where $\Omega_n\equiv n\Omega$. Below, we also write $k=|\mathbf{k}|$. For a fixed positive radiative harmonic of a Keplerian binary, $\Omega_n a=\mathcal{O}(v)$. The on-shell wave number $k_n=\sqrt{\Omega_n^2-m^2}\leq\Omega_n$, so $k_na=\mathcal{O}(v)$ away from the mass threshold and is further suppressed near the threshold. In the following, we assume that the radiation is sufficiently relativistic.\footnote{Near the radiation threshold, PN power counting and the far-zone asymptotic expansion may require a more careful treatment.} Expanding $ e^{-\ii\mathbf{k}\cdot\mathbf{X}_I}$ and $1/\gamma_I$ through $\mathcal{O}(v^2)$ gives $n_{(n)}(\mathbf{k})
=-\ii k^iD_n^i-k^i k^j I_n^{ij}-K_n+\mathcal{O}(v^3)$, where
\begin{align}
 D_n^i&=\frac{1}{T}\int_0^Tdt\,e^{\ii\Omega_n t}
 \sum_I q_{I\phi}X_I^i,
 \\
 I_n^{ij}&=\frac{1}{2T}\int_0^Tdt\,e^{\ii\Omega_n t}
 \sum_I q_{I\phi}X_I^iX_I^j,
 \\
 K_n&=\frac{1}{2T}\int_0^Tdt\,e^{\ii\Omega_n t}
 \sum_I q_{I\phi}\mathbf v_I^2.
\end{align}
Away from the radiation threshold, the quadratic term in the expansion and the monopole term $-K_n$ are both $\mathcal{O}(v^2)$ and enter at the ``quadrupole order,'' namely, the order in the long-wavelength expansion at which the angular $l=2$ multipole first appears.

At quadrupole order, however, the Cartesian moments appearing in the Taylor expansion are not themselves irreducible angular multipoles. A symmetric Cartesian moment of rank $p$ is, in general, reducible under rotations; its decomposition into STF tensors contains irreducible components of ranks $l=p,p-2,\ldots$, with successive traces determining the lower-rank components~\cite{Ross2012}. For the second moment, write $I_n^{ij}=Q_n^{ij}+\frac{1}{3}\delta^{ij}I_n$, $Q_n^{ii}=0$, $I_n=I_n^{ii}$. With $k^i=kn_i$, and $n_{\langle ij\rangle}\equiv n_in_j-\delta_{ij}/3$, $ -k^i k^jI_n^{ij}
=-k^2 n_{\langle ij\rangle}Q_n^{ij}
-\frac{k^2}{3}I_n$. Thus, the STF part $Q_n^{ij}$ contributes to the angular $l=2$ component, whereas the trace $I_n$ contributes to the $l=0$ component. The source through $\mathcal O(v^2)$ can therefore be organized into
\begin{equation}
 n_{(n)}(k\mathbf n)
 =\mathcal A_n^{(0)}
 +n_i\mathcal A_n^{(1)i}
 +n_{\langle ij\rangle}\mathcal A_n^{(2)ij}
 +\mathcal O(v^3),
 \label{erratum:eq:angular-decomposition}
\end{equation}
with $ \mathcal A_n^{(0)}=-K_n-\frac{k^2}{3}I_n$, $
\mathcal A_n^{(1)i}
=-\ii kD_n^i$, and $
\mathcal A_n^{(2)ij}
=-k^2Q_n^{ij}$. Note that Ref.~\cite{ChengWuCao2025} refers to the term $-k^i k^jI_n^{ij}$ as ``scalar quadrupole radiation.'' In the binary center-of-mass frame, a circular orbit has constant $\sum_I q_{I\phi}X_I^2$ and $\sum_I q_{I\phi}\mathbf v_I^2$, so $I_n=K_n=0$ for $n\neq0$; for a noncircular orbit these quantities are time dependent. The multipole expansion of Ref.~\cite{KrauseKloorFischbach1994} likewise does not separate the irreducible angular multipoles.

The $\mathcal{O}(v^2)$ amplitudes displayed above do not constitute a complete radiation calculation at that accuracy in full GR. When the leading $\mathcal{O}(v)$ dipole amplitude is nonzero, a scalar energy flux complete through relative 1PN order requires the complete $\mathcal{O}(v^2)$ monopole and quadrupole sectors, together with the $\mathcal{O}(v^3)$ correction to the dipole~\cite{KuntzPiazzaVernizzi2019}. The $\mathcal{O}(v^3)$ dipole amplitude correction interferes with the leading dipole amplitude at the same order as the squared $\mathcal{O}(v^2)$ monopole and quadrupole amplitudes. If the leading dipole vanishes, the complete $\mathcal{O}(v^2)$ $l=0$ and $l=2$ sectors instead determine the leading radiation, with the monopole sector receiving additional gravitational contributions at the same order in full GR [at this order, the factor $1-\frac{|\mathbf v_I|^2}{2}$ in Eq.~\eqref{erratum:eq:source-flat-pn} is replaced by $1-\frac{|\mathbf v_I|^2}{2}+U_I$, where $U_I$ is the Newtonian potential sourced by the other bodies evaluated at $\mathbf{X}_I$], which were omitted in Ref.~\cite{ChengWuCao2025}. The corresponding radiation fluxes for eccentric Keplerian binaries with scalar charges in a massive scalar theory have yet to be calculated. Such additional gravitational contributions are absent at this order in the Proca case, and the corresponding energy flux has been calculated in Ref.~\cite{ChengWuCao2025}.

\section*{III. Flux checks and correction to the Proca angular-momentum flux}

The scalar and Proca flux formulas derived in this paper are not restricted to a globally flat spacetime, so long as the source is periodic (with respect to the asymptotic inertial time). For an asymptotically flat solution of the full nonlinear Einstein-matter equations in the same minimally coupled scalar or Proca theory, the leading scalar or Proca flux is determined by the same quadratic functional of the asymptotic radiation field. In the wave zone, $g_{ab}=\eta_{ab}+\mathcal{O}(1/r)$ while the radiative scalar or Proca field is $\mathcal{O}(1/r)$, so metric corrections to the flux formulas are subleading. What changes beyond the flat-spacetime approximation is the relation between the near-zone source and the asymptotic radiative amplitudes. The point-particle Fourier amplitudes in flat spacetime are replaced by effective radiative amplitudes that incorporate gravitational dressing. In the PN description, the additional gravitational contribution to the scalar monopole amplitude discussed above is one example of the nonlinear near-zone corrections that enter the effective radiative source after the potential modes are integrated out~\cite{KuntzPiazzaVernizzi2019}.

Besides the derivation in this paper, the flux formulas can be independently checked in flat spacetime from the rates of energy, linear-momentum, and angular-momentum transfer between the retarded field and the periodic source. This provides a cross-check of the formulas. Throughout this section, the angular momentum is defined about the spatial origin associated with the coordinates $\mathbf{x}$. We define $\avg{Q}\equiv\frac{1}{T}\int_0^Tdt\,Q(t)$ and
$G_R(\omega,k)\equiv\frac{1}{(\omega+\ii\,0^+)^2-k^2-m^2}$.
The absorptive part of the retarded propagator is
\begin{equation*}
 \Gamma(\omega,k)\equiv-2\,\mathrm{Im}\,G_R(\omega,k)
 =2\pi\,\mathrm{sgn}(\omega)\delta(\omega^2-k^2-m^2).
\end{equation*}

\subsection*{A. Klein-Gordon field}

For $(\square-m^2)\phi=-n$, the scalar energy-momentum tensor obeys $\partial_aT_\phi^{ab}=n\,\partial^b\phi$. Let $\phi_\texttt{R}$ denote the retarded field. The energy-momentum balance and the moment of the force density $f^a_\texttt{R}=-n\,\partial^a\phi_\texttt{R}$ give the source-side relations for the dissipation (radiation) fluxes:
\begin{align*}
	P&
	=-\avg{\int d^3x\,f^0_\texttt{R}}=\avg{\int d^3x\,n\,\partial_t\phi_\texttt{R}},\\
	F_i&
	=-\avg{\int d^3x\,f^i_\texttt{R}}=-\avg{\int d^3x\,n\,\partial_i\phi_\texttt{R}},\\
	\tau_i&=-\avg{\int d^3x\,\eps_{ijk}x^j f^k_\texttt{R}}
	=-\avg{\int d^3x\,n\,\eps_{ijk}x^j\partial_k\phi_\texttt{R}}.
\end{align*}
Expanding the retarded field in Fourier modes, with $\phi_{(n)}(\mathbf{k})\equiv T^{-1}\int_0^Tdt\,e^{\ii\Omega_n t}\int d^3x\,e^{-\ii\mathbf{k}\cdot\mathbf{x}}\phi_\texttt{R}(t,\mathbf{x})$, gives $\phi_{(n)}(\mathbf{k})=-G_R(\Omega_n,k)\,n_{(n)}(\mathbf{k})$. The time average pairs the positive- and negative-frequency harmonics, whose principal-value contributions cancel, leaving only the absorptive kernel $\Gamma_n\equiv\Gamma(\Omega_n,k)$. The results are
\begin{align}
	P&=\sum_{n>0}\Omega_n\intk
	\Gamma_n\left|n_{(n)}\right|^2,
	\label{erratum:eq:scalar-source-P}\\
	F_i&=\sum_{n>0}\intk
	\Gamma_n k^i\left|n_{(n)}\right|^2,
	\label{erratum:eq:scalar-source-F}\\
	\tau_i&=\sum_{n>0}\intk
	\Gamma_n\,
	\mathrm{Re}\!\left[
	n_{(n)}^*\,(\widehat L_{\mathbf{k}})_i\, n_{(n)}
	\right],
	\label{erratum:eq:scalar-source-L}
\end{align}
where $(\widehat L_{\mathbf{k}})_i=-\ii\,\eps_{ijk}k^j\frac{\partial}{\partial k^k}$. Performing the radial $k$-integral over radiative harmonics, for which $\Omega_n>m$ and $k_n=\sqrt{\Omega_n^2-m^2}$, reproduces Eqs.~(19), (21), and (17), respectively. For a prescribed classical source and vacuum initial state, the on-shell amplitudes $n_{(n)}$ also determine the coherent state of the emitted field, whose mean emission rates of the corresponding Noether charges reproduce the time-averaged radiation fluxes, which can equivalently be organized in terms of irreducible angular multipoles.

\subsection*{B. Proca field}
In the Proca case, let $A_a^\texttt{R}$ be the retarded field, $F_{ab}^\texttt{R}=\partial_aA_b^\texttt{R}-\partial_bA_a^\texttt{R}$, and $f^a_\texttt{R}= -\partial_bT_A^{ba}=J_bF^{\texttt{R}ba}$. We have
\begin{align*}
 P&=-\avg{\int d^3x\,f^0_\texttt{R}}
 =\avg{\int d^3x\, J^aF^{\texttt{R}}_{0a} },\\
 F_i&=-\avg{\int d^3x\,f^i_\texttt{R}}
 =\avg{\int d^3x\,J^aF^{\texttt{R}}_{ai} },\\
 \tau_i&=-\avg{\int d^3x\,\eps_{ijk}x^j f^k_\texttt{R}}=\avg{\int d^3x\,\eps_{ijk}x^j J^aF^{\texttt{R}}_{ak} }
 \\
 &=-\left\langle\int d^3x\,
 \epsilon_{ijk}x^jJ^a\partial_kA_a^\texttt{R}\right\rangle
 -\left\langle\int d^3x\,
 \epsilon_{ijk}J^jA_k^\texttt{R}\right\rangle
 ,
\end{align*}
where in the last line we used $\partial_a(x^jJ^aA_k^\texttt{R})
=(\partial_ax^j)J^aA_k^\texttt{R}
+x^j(\partial_aJ^a)A_k^\texttt{R}
+x^jJ^a\partial_aA_k^\texttt{R}$. Inserting the Fourier-space retarded solution $A^a(\mathbf{k})=G_R(\Omega_n,k)\,J^a_{(n)}(\mathbf{k})$, and pairing the positive- and negative-frequency harmonics again leaves only the absorptive kernel $\Gamma_n$. Current conservation gives $\Omega_nJ_{(n)}^0(\mathbf{k})=k^iJ_{(n)}^i(\mathbf{k})$. We obtain
\begin{align}
	P&=\sum_{n>0}\Omega_n\intk\Gamma_n\mathcal W_n,
	\label{erratum:eq1}\\
	F_i&=\sum_{n>0}\intk\Gamma_n k^i\mathcal W_n,
	\label{erratum:eq2}\\
	\tau_i&=\sum_{n>0}\intk\Gamma_n\,
	\mathrm{Re}\,\mathcal{V}_n^i,\label{erratum:eq3}
\end{align}
where $\mathcal W_n
\equiv
-J^*_{(n)a}J_{(n)}^a=
J_{(n)}^{i*}J_{(n)}^i
-\frac{k^i k^j}{\Omega_n^2}
J_{(n)}^{i*}J_{(n)}^j$, and $\mathcal V_n^i
\equiv
-J_{(n)a}^*(\widehat L_{\mathbf{k}})_iJ_{(n)}^a
-\ii\,\eps_{ijk}J_{(n)}^{j*}J_{(n)}^k$. After the radial $k$-integral, Eqs.~\eqref{erratum:eq1} and \eqref{erratum:eq2} reproduce Eqs.~(28) and (30). Equation~\eqref{erratum:eq3} instead gives the corrected angular-momentum flux.

To identify the two corrections in the original derivation, we return to the second line of Eq.~(25) and use the same definitions $D_l\equiv\partial_l+\ii k_n n_l$ and $D_l^*\equiv\partial_l-\ii k_n n_l$. The first term should read $-\ii\Omega_nJ_{(n)}^{j*}D_lJ_{(n)}^j$, rather than $+\ii\Omega_nJ_{(n)}^{j*}D_lJ_{(n)}^j$. In addition, the term that was dropped in passing to the third line expands as
\begin{align}
 -D_jJ_{(n)}^0D_l^*J_{(n)}^{j*}
 ={}&-\left[\partial_jJ_{(n)}^0\right]\left[\partial_lJ_{(n)}^{j*}\right]
 +\ii k_n n_lJ_{(n)}^{j*}\partial_jJ_{(n)}^0
 -\ii k_n n_jJ_{(n)}^0\partial_lJ_{(n)}^{j*}
 -k_n^2n_jn_lJ_{(n)}^0J_{(n)}^{j*}.
 \label{erratum:eq:omitted-proca-expansion}
\end{align}
For a far-zone amplitude that depends on position only through the direction $\mathbf n=\mathbf x/r$, one has $r\partial_l =\left(\delta_{la}-n_ln_a\right)\frac{\partial}{\partial n_a}$. After multiplication by the prefactor $\epsilon_{ikl}\,r\,n_k$ in Eq.~(25), the first term in Eq.~\eqref{erratum:eq:omitted-proca-expansion} vanishes as $r^{-1}$, while the second and fourth terms vanish because they are proportional to $\epsilon_{ikl}n_kn_l$.  The third term instead survives at the same far-zone order as the retained terms and therefore cannot be discarded.

Thus, two corrections are required. First, the first term in Eq.~(26) changes sign. Second, the third term of Eq.~\eqref{erratum:eq:omitted-proca-expansion} must be retained. In the positive-frequency convention used there, the latter correction consists of adding
\begin{align*}
 \Delta\tau_i={}&\sum_{n\ge n_0=\frac{m}{\Omega}}
 \frac{k_n^2}{8\pi^2\Omega_n}
 \int d\Omega_{\mathbf n}\,\mathrm{Re}\!\left[
 -\ii\eps_{ikl}r n_kn_jJ_{(n)}^0\partial_lJ_{(n)}^{j*}
 \right]
\end{align*}
to the right-hand side of Eq.~(26). Here, $\frac{k_n^2}{8\pi^2\Omega_n}=\frac{1}{16\pi^2}\times k_n\times v_g\times 2$, where $k_n$ comes from the third term of Eq.~\eqref{erratum:eq:omitted-proca-expansion}, $v_g=k_n/\Omega_n$ is the group velocity, and the factor of $2$ comes from pairing the positive- and negative-frequency harmonics.

The correction does not affect the dipole result used later in this paper.  In the dipole approximation of Eqs.~(31) and (32), $J_{(n)}^i=\Omega_n j_n^i$ is independent of the radiation direction and $J_{(n)}^0=k_n n_i j_n^i$.  Hence $\partial_lJ_{(n)}^i=0$, so both the sign-corrected term and the added term $\Delta\tau_i$ vanish. The remaining terms therefore give Eq.~(35).

\nocite{*}
\begingroup
\let\erratumBibLabel\label
\renewcommand{\label}[1]{\erratumBibLabel{erratum:#1}}
\putbib[erratum_refs]
\endgroup
\end{bibunit}
\endgroup

\end{document}